\documentclass[%
 aip,
 amsmath,
amssymb,
 reprint,%
]{revtex4-1}

\usepackage{graphicx}
\usepackage{dcolumn}
\usepackage{bm}

\usepackage[utf8]{inputenc}
\usepackage[T1]{fontenc}
\usepackage{mathptmx}
\usepackage{etoolbox}

\usepackage{paralist}
\DeclareMathAlphabet{\mathcal}{OMS}{cmsy}{m}{n}
\DeclareSymbolFont{largesymbols}{OMX}{cmex}{m}{n}

\makeatletter
\def\@email#1#2{%
 \endgroup
 \patchcmd{\titleblock@produce}
  {\frontmatter@RRAPformat}
  {\frontmatter@RRAPformat{\produce@RRAP{*#1\href{mailto:#2}{#2}}}\frontmatter@RRAPformat}
  {}{}
}%
\makeatother
\begin{document}

\preprint{AIP/123-QED}

\title{On the singularity of Lie-transform perturbation approach to the guiding-center problem}
\author{W. H. Lin}
\affiliation{Southwestern Institute of Physics, Chengdu 610041, China.}
\author{J. Garcia}%
\affiliation{ 
CEA, IRFM, F-13108 Saint Paul-lez-Durance, France.
}%

\author{J. Q. Li}
\email{lijq@swip.ac.cn}
\affiliation{%
Southwestern Institute of Physics, Chengdu 610041, China.
}%

\date{\today}

\begin{abstract}
We present a novel scheme of carrying out the Lie-transform perturbation for the guiding-center motion, with an aim at addressing directly the problem of singularity which exists intrinsically in the determining equation for the generating vector, and which gives rise to the formidable gauge functions in the pure oscillating part of the Lie transformation. Whereas in most applications of Lie-transform perturbation such gauge functions must be approximately solved from some partial differential equations, our scheme, characterized by a staggered determination of the generating vectors, naturally produces the gauge functions through explicit integral over the gyro-angle, leaving no unaccountable error of high order in all the succeeding transformations. Based on such scheme, a formalism of guiding-center transformation has been derived in a unified manner retaining the effects of the strong $E\times B$ shearing as well as those of electromagnetic fluctuations. 
\end{abstract}

\maketitle

\section{\label{sec:Introduction}Introduction}
Guiding-center motion is a classical problem of fundamental importance to plasmas physics. Small parameters such as the ratio of the Larmour radius to the characteristic length of the strong magnetic field make the perturbation analysis possible, and in the past decades a variety of perturbation theories\cite{kruskal1962asymptotic,northrop1963adiabatic,littlejohn1981hf,littlejohn1983variational,weyssow1986hamil} had been developed, among which the Hamiltonian approaches (see Ref.~\onlinecite{cary2009hamiltonian} for a review) had gained their popularity owing to the availability of conservation laws such as that of energy and of phase space volume.

The Hamiltonian Lie-transform perturbation (HLTP) approach developed in Ref.~\onlinecite{littlejohn1982hamiltonian}, not to be confused with other perturbation approaches lumped together in the name of Lie transform\cite{cary1981lie},  deals directly with the Poincar\'e-Cartan One-Form (PCOF) without referring to any specific coordinates, and thus produces a systematic procedure whereby the task of performing perturbation transformation is reduced to the pure labor of calculating algebraically the generating vectors. Adopting the so-called active point of view, the basic strategy of the HLTP approach in Ref.~\onlinecite{littlejohn1982hamiltonian} is not to seek a complete set of new coordinates -- often called guiding-center coordinates as opposed to the particle coordinates -- in which the new representation of the original PCOF is independent on the gyro-angle; but rather, it is to transform the original PCOF under the flow of a series of generating vectors into a geometrically different PCOF that is by construction symmetric along the gyro-angle.

The purpose of the present study is to reveal, for the guiding-center motion as well as for the general perturbation motion, the singularity of HLTP transformation which resides in the inherent singularity of any two-form on the odd-dimensional phase space where the Hamiltonian function is treated as a mere component of the PCOF. The generating vector of HLTP transformation, as a consequence, cannot be obtained from a non-singular equation of linear algebra without solving at first the gauge function from a partial differential equation, which could well be formidable if the unperturbed counterpart of the perturbation motion is non-trivial in its own right. In regard to the two-step HLTP scheme\cite{brizard2007foundations} where the spatial non-uniformity of the equilibrium fields and the low-frequency fluctuation of the perturbed ones are treated separately, the solution of such gauge function is approximated from an indefinite integral over the gyro-angle, leaving an error of high order in the generating vector that cannot be accounted for in the succeeding transformation. As interests of high-order gyrokinetic formalism\cite{qin2007geometric,hahm2009fully,parra2011phase,burby2013auto,tronko2015lagrangian,brizard2017variational} have grown recently in connection with the concern that the fluctuations of relatively high amplitude in the edge region of tokamak plasmas may affect the accuracy of gyrokinetic simulations, the error within the gauge function, though inconsequential for the gyrokinetic Vlasov equation, may pose a potential obstacle in obtaining the pull-back transformations of the desired accuracy.

The present study is then dedicated to developing a novel scheme of HLTP transformation where the error within gauge function does not exist in the first place as its determining equation will be an integrable one. This scheme featuring a staggered determination of each generating vector derives from our general consideration on a less considered type of perturbation motion, which we propose to call the singular perturbation motion, for it does not naturally approach to a well-defined unperturbed counterpart as the small parameter characterizing the strength of perturbation approaches to zero. We show that, after some extra conditions are properly introduced, the HLTP method to singular perturbation motion is not less sound and workable than to the non-singular ones, and that a unified HLTP treatment of the perturbation effects which arise separately from equilibrium and perturbed fields, can be accomplished for the guiding-center motion, were it to be interpreted as a special case of singular perturbation motion. Thus, we derived, including consistently the effects of the strong $E \times B$ shearing which plays a important role in regulating the transports in tokamak plasmas\cite{connor2004itb}, a formalism of guiding-center transformation that we believe is suitable to the full-f gyrokinetic simulations.

The paper is organized as follows. In Sec.~\ref{sec:Preliminaries}, the geometric foundation of HLTP is briefly introduced with an explanation of the assumed notations and conventions. In Sec.~\ref{sec:general} we deal with the application of HLTP method to general perturbation motion, the results of which are used in Sec.~\ref{sec:gc_trans} in dealing with the guiding-center motion in a dominant equilibrium magnetic and electric field. Finally, we make our conclusions and discussions in Sec.~\ref{sec:concl}.

\section{\label{sec:Preliminaries}Preliminaries}

The terminologies of differential geometry will be frequently involved; readers who find them unfamiliar may refer to, for example, Ref.~\onlinecite{arnold1989mathe,chern1999lectures}. Let $\mathcal{C}$ be the three dimensional configuration space, $T \mathcal{C}$ and $T^* \mathcal{C}$ be its tangent and cotangent bundle respectively, and the extended phase pace $\mathcal{M}$ be the manifold product of $T^* \mathcal{C}$ and real number space $R$, i.e. $\mathcal{M} = R \times T^* \mathcal{C}$. In the odd-dimensional space $\mathcal{M}$, particle motions are generated by the vortex vector $X$ of an one-form $\vartheta \in T^* \mathcal{M}$ that satisfies,
\begin{eqnarray} i_{X} d \vartheta = 0 \label{eq:DF_vortex},\end{eqnarray}
where $d$ is the exterior derivative and $i_{Y}$ for arbitrary vector $Y \in T \mathcal{M}$ is a linear operator mapping $n$-forms to $(n-1)$-forms. Such $X$ is also called a null vector of the two-form $d \vartheta$. Once a set of functions $\{u^{\alpha}; 1 \le \alpha \le 6\}$ together with $u^0 \equiv t$, the time coordinate, is selected as a coordinate system in some subset of $\mathcal{M}$, their exterior derivatives $\{d u^{\alpha}\}$ serve as a natural basis of $T^* \mathcal{M}$, and a dual basis $\{\partial / \partial u^{\alpha}\}$ of $T \mathcal{M}$ can be accordingly defined by the relations, for $0 \le \alpha, \beta \le 6$
\begin{eqnarray} 
   \left\langle \frac{\partial}{\partial u^{\alpha} }, d u^{\beta} \right\rangle = \delta^{\beta}_{\alpha}
\label{eq:DF_dual_base},
\end{eqnarray}
where $\delta$ is the Kronecker delta symbol and $\langle , \rangle$ is a bilinear, point-wise bracket which maps the pairing of contravariant and covariant tensors at a point of $\mathcal{M}$ to a real number (and therefore maps the pairing of tensor fields to a real function on $\mathcal{M}$). The collection of vectors $\{ \partial / \partial u^{\alpha}\}$ is called the \emph{natural basis} of the coordinates $\{ u^{\alpha}\}$. Eq.~(\ref{eq:DF_vortex}) is then represented by their components $X^{\alpha} \equiv \langle X, du^{\alpha} \rangle$ and $\vartheta_{\beta} \equiv \langle \frac{\partial}{\partial u^{\beta} }, \vartheta \rangle$, 
\begin{eqnarray} 
X^{\alpha} \left(\frac{\partial \vartheta_{\beta}}{\partial u^{\alpha}} - \frac{\partial \vartheta_{\alpha}}{\partial u^{\beta}} \right ) = 0
\label{eq:vortex_component},
\end{eqnarray}
where summation convention has been used and will be used throughout the paper unless otherwise mentioned. The existence of non-zero $X$ is guaranteed by the simple fact that an anti-symmetric odd-dimensional matrix, such as defined by $\omega_{\alpha \beta} \equiv \partial \vartheta_{\alpha} / \partial u^{\beta} - \partial \vartheta_{\beta} / \partial u^{\alpha}$, must be singular (having zero determinant). But apparently such $X$ is not unique. We will exclusively call an one-form a Poincar\'e-Cartan One-Form (PCOF) if it has exactly one linearly independent vortex vector, so that two vectors both satisfying Eq.~(\ref{eq:DF_vortex}) must differ from one another merely by the multiplication of some scalar function, which is often used to scale to unity the component of $X$ along the time coordinate. Therefore, a PCOF with its vortex vector defines a certain type of particle dynamic whose motion equation in coordinates $\{ u^{\alpha}\}$ is given by
\begin{eqnarray} 
\frac{d u^{\alpha}}{dt} \equiv \langle X, du^{\alpha} \rangle
\label{eq:motion}. 
\end{eqnarray}
As to the PCOF dependent on some small parameter $\epsilon$, we call the associated particle motion a non-singular perturbation motion if the PCOF with $\epsilon$ remains a PCOF when $\epsilon$ approaches to zero, otherwise it is called the singular perturbation motion; the former has a well-defined unperturbed counterpart directly obtained by setting $\epsilon = 0$, while the vortex vector of the latter typically has a leading term of order $o(\epsilon^{-1})$ after the time component is scaled to $o(\epsilon^0)$. It should be noted in particular that an equivalent description of particle motion can be established in the even-dimensional phase space $T^* \mathcal{C}$ with the time coordinate treated as a pure parameter. The particle motions are depicted therein as following the flow of Hamiltonian vector which is uniquely associated with an Hamiltonian function by the symplectic structure of even-dimensional phase space. But in this work we will make use of only the description in odd-dimensional phase space $\mathcal{M}$. 

Of direct relevance to HLTP approach is the induced particle dynamic of the self-diffeomorphism $\mathcal{F}: \mathcal{M} \rightarrow \mathcal{M}$. Under its induction maps $\mathcal{F}_{*}$ and $\mathcal{F}^{*}$, the tensor fields on $\mathcal{M}$ are transformed satisfying the relation,
\begin{eqnarray} 
\langle Y_1 \wedge Y_2 \cdots \wedge Y_k , \mathcal{F}^{*} \omega \rangle = \langle \mathcal{F}_*  Y_1 \wedge \mathcal{F}_* Y_2 \cdots \wedge \mathcal{F}_* Y_k, \omega \rangle
\label{eq:induced_bracket},
\end{eqnarray}
where $Y_1, Y_2, \cdots , Y_k \in T \mathcal{M}$, $\omega$ is a $k$-form and $\wedge$ is the exterior (or wedge) product which commutes with the induction maps, satisfying $\mathcal{F}_* (Y_1 \wedge Y_2) = \mathcal{F}_*  Y_1 \wedge \mathcal{F}_* Y_2$ for two arbitrary vectors and so on. Note that in Eq.~(\ref{eq:induced_bracket}) we have suppressed, for convenience, the subscripts denoting where the brackets are evaluated, and that no ambiguity will thus arise as Eq.~(\ref{eq:induced_bracket}) is only invoked in this work when both sides are everywhere zero (or equal to negligible small terms). Let $\bar{\vartheta}$ denote the image of $\vartheta$ under the induction map $\mathcal{F}^*$, namely, $\bar{\vartheta} \equiv \mathcal{F}^* \vartheta$. Given that $\mathcal{F}$ is a diffeomorphism, it is easy to prove (\romannumeral1) that $\bar{\vartheta}$ is a PCOF if and only if $\vartheta$ is itself a  PCOF, and (\romannumeral2) that the vortex vectors of $\vartheta$ and $\bar{\vartheta}$, denoted respectively by $X$ and $\bar{X}$, are related to each other by the relation $X = \mathcal{F}_* \bar{X}$. As long as $\mathcal{F}$ is not an identity map, $\bar{\vartheta}$ is generally different from $\vartheta$, and we therefore say there is an induced particle dynamic associated with $\bar{\vartheta}$, which is often non-physical and different from what $\vartheta$ has defined when both observed in the same coordinate. The motion equations of these two particle dynamic are related to one another by Eq.~(\ref{eq:induced_bracket}) for $k=1$, i.e.,
\begin{eqnarray} 
\langle X , d \mu \rangle = \langle \bar{X} , d \mathcal{F}^* \mu \rangle
\label{eq:induced_dynamic_3},
\end{eqnarray}
where $\mu$ is arbitrary scalar function on $\mathcal{M}$ and can be typically substituted with the coordinate functions $\{ u^{\alpha} \}$. It is implied in Eq.~(\ref{eq:induced_dynamic_3}) that the solution $\mu$ to the equation $\langle X , d \mu \rangle = 0$ can be obtained by solving at first $\langle \bar{X} , d \bar{\mu} \rangle = 0$, which could be easier to solve than the former if $\mathcal{F}$ is properly constructed, and secondly calculating $\mu = (\mathcal{F}^{-1})^* \bar{\mu}$. 

Another mathematical fact of relevance is that a smooth vector field $G$ on $\mathcal{M}$ generates locally a one-parameter group of self-diffeomorphisms, denoted by $\mathcal{F}^{s}$ and characterized by the properties that (\romannumeral1) $\mathcal{F}^{0}$ is an identity map and (\romannumeral2) $\mathcal{F}^{r+s} = \mathcal{F}^{r}\mathcal{F}^{s}$ (and therefore $(\mathcal{F}^{s})^{-1} = \mathcal{F}^{-s})$, where $r$ and $s$ are small real numbers. An oft-used synonym of this $\mathcal{F}^{s}$ is the \emph{phase flow} (or simply \emph{flow}) generated by $G$. Let $\Phi^{s}$ be such an united map that $\Phi^s \sigma = (\mathcal{F}^s)^* \sigma$ when $\sigma$ is a covariant tensor and $\Phi^s \sigma = (\mathcal{F}^{-s})_* \sigma$ when $\sigma$ contravariant. For small $s$, $\Phi^s$ is a near-identity map, and the Lie derivative $L_G$ is defined to measure the rate of change of a tensor field $\sigma$ under $\Phi^s$, 
\begin{eqnarray} 
\Phi^s \sigma - \sigma = s L_G \sigma + o(s^2) 
\label{eq:DF_Lie_derivative}.
\end{eqnarray}
Quite naturally, the tensor field $\sigma$ satisfying $L_G \sigma=0$ is called an invariant of the flow generated by $G$. A corollary to the definition of $L_G$ is that $d\Phi^s / ds = \Phi^s L_G$, with which we can obtain a formal expansion of $\Phi^s$ for small s,
\begin{eqnarray} 
\Phi^s = 1 + s L_G + \frac{1}{2} s^2 L_G L_G + \cdots \equiv exp(s L_G) 
\label{eq:exp_LG}.
\end{eqnarray}
The relations between the operators, $L_Y$, $d$ and $i_Z$ for arbitrary vectors $Y$ and $Z$ are given by the Cartan formulas, among which the following two prove to be most useful,
\begin{eqnarray} 
L_Y = i_Y d + d i_Y \label{eq:cartan_formulas_1},\\
L_Y i_Z - i_Z L_Y = i_{[Y,Z]}
\label{eq:cartan_formulas_2},
\end{eqnarray}
where $[Y,Z]\equiv YZ - ZY$ is the Poisson bracket of vectors. No doubt the vortex vector $X$ itself determines a group of self-diffeomorphisms $\mathcal{F}^s_X$, often called Hamiltonian phase flow, which preserves the two-form $d \vartheta$ and hence a series of $2n$-form $d \vartheta \wedge d \vartheta \cdots \wedge d\vartheta$.

\section{\label{sec:general}general structure of the HLTP transformation}

Suppose the perturbed system in question is determined by a PCOF $\Theta$ in its infinite power series of $\epsilon$,
\begin{eqnarray} 
\Theta \equiv dS + \vartheta \equiv dS + \vartheta_0 + \epsilon^1 \vartheta_{1} + \cdots + \epsilon^n \vartheta_{n} + \cdots
\label{eq:DF_Theta}.
\end{eqnarray}
No assumption is needed about whether the $\vartheta_n$ for $n > 0$ can be itself in a power series of $\epsilon$ or whether they can depend on other small parameters. But we assume (\romannumeral1) that all the quantities that don't depend on $\epsilon$ explicitly are of $o(\epsilon^0)$, (\romannumeral2) that the components of $\Theta$ is known to us under the coordinates $\{u^{\alpha}, 0 \le \alpha \le 6\}$ among which $u^{0} \equiv t$ is the time and $u^{6} \equiv \xi$ is an angle-like variable, and (\romannumeral3) that the unperturbed term $\vartheta_0$ is an invariant of the vector $\zeta \equiv \partial / \partial \xi$ with the condition
\begin{eqnarray} 
L_{\zeta} \vartheta_0 = 0
\label{eq: symmetric_1}.
\end{eqnarray}
If the same condition were to be realized for the whole $\vartheta$ up to arbitrary order of $\epsilon$ (i.e. $L_{\zeta}\vartheta=0$), it is easy to prove with Eq.(\ref{eq:cartan_formulas_2}) that the vortex vector $X$ of $\vartheta$ must be an invariant of the flow generated by $\zeta$, namely, that 
\begin{eqnarray} 
L_{\zeta} X = [\zeta,X]=0
\label{eq: symmetric_2}.
\end{eqnarray}
A dynamic system is then established which possesses a symmetry along $\zeta$ and accordingly a constant of motion that is exactly $i_{\zeta} \vartheta$. As an aside, it may appear that our criterion of symmetry, $L_{\zeta} \vartheta=0$, is more strict and therefore less general than what has been used in Ref.~\onlinecite{qin2007geometric},
\begin{eqnarray} 
 L_{\zeta} \vartheta'=d S'
\label{eq: symmetric_3},
\end{eqnarray}
but this is not the case considering the fact that, whenever $\vartheta '$ satisfies Eq.(\ref{eq: symmetric_3}), it is always possible to separate in advance an exact term $dS$ (with $S$ defined by $i_{\zeta}dS = S '$) from $\vartheta '$ so that the remaining term will immediately satisfy our criterion; this explains the appearance of $dS$ in Eq.(\ref{eq:DF_Theta}).

However, such symmetry often ceases to exist upon including the very first order term $\vartheta_1$. According to the above discussion, this is equivalent to say that
\begin{eqnarray} 
L_{\zeta} d \Theta  = o(\epsilon^1).
\label{eq: sym_0th}
\end{eqnarray}
A well-established method to deal with such asymmetry along $\zeta$ is to seek a complete set of new coordinates whose natural basis contain a vector, usually the first-order modification of $\zeta$, along which $d \Theta$ is symmetric up to arbitrary order. Such method seeking to transform the coordinates instead of $\Theta$ itself is called the perturbation method in the passive point of view\cite{littlejohn1982hamiltonian}, as compared with the active point of view in which the $\Theta$ must be transformed into a geometrically new PCOF. The active point of view is more straightforward to us and thus adopted throughout the paper. (But it can be easily proved that this two point of view are equivalent in terms that, whenever the transformation meeting our requirements is worked out in one of them, a corresponding transformation is immediately obtained in the other). Following the methodology of HLTP, we seek a series of groups of self-diffeomorphisms, $\mathcal{F}^s_1,\mathcal{F}^s_2,\cdots,\mathcal{F}^s_N$, generated respectively by the vectors $G_1,G_2,\cdots,G_N$, and, with the element $s=-\epsilon^n$ of each group $\mathcal{F}^s_n$, $1 \le n \le N$, construct a composite self-diffeomorphism,
\begin{eqnarray} 
\mathcal{F} \equiv \mathcal{F}_1^{-\epsilon^1} \circ \cdots \circ \mathcal{F}_N^{-\epsilon^N},
\label{eq:comps_diffmorp}
\end{eqnarray}
whose induction map $\mathcal{F} ^*$, 
\begin{eqnarray} 
\mathcal{F} ^* = exp(-\epsilon^N L_{G_N}) \cdots exp(-\epsilon L_{G_1}),
\end{eqnarray}
transforms the original $\Theta$ into a new PCOF, $\Theta^{(N)} = \mathcal{F}^* \Theta$, which is constructed with the property,
\begin{eqnarray} 
L_{\zeta} d \Theta^{(N)}  = o(\epsilon^{N+1}).
\label {eq: sym_nth}
\end{eqnarray}
Before we consider the questions as to how the new PCOF $\Theta^{(N)}$ can be constructed and subsequently how the generating vectors $G_n$'s are determined, it proves useful to define a series of intermediate PCOFs by the recursive relation
\begin{eqnarray} 
\Theta^{(n)} \equiv exp(-\epsilon^n L_{G_n}) \Theta^{(n-1)}.
\label {eq:intermediate_PCOFs}
\end{eqnarray}
Both the $\Theta^{(n)}$ and $\Theta^{(n-1)}$ can be cast into a expansion series of $\epsilon$ similar to Eq.~(\ref{eq:DF_Theta}) but with the superscript of the bracketed number, i.e.,
\begin{eqnarray} 
\Theta^{(n)} \equiv dS^{(n)} + \vartheta^{(n)} \equiv  dS^{(n)}  + \sum_{m=0}^{\infty} \epsilon^m \vartheta^{(n)}_m,
\label{eq:Theta_n}
\end{eqnarray}
for $1 \le n \le N$. The term $d S^{(n)}$ is defined to collect all the exact terms obtained from invoking repeatedly Eq.~(\ref{eq:cartan_formulas_1}), while the term $\vartheta^{(n)}$ is to collect the remaining:
\begin{subequations}
\begin{equation}
\vartheta^{(n)} \equiv exp(-\epsilon^n i_{G_n} d) \vartheta^{(n-1)},
\end{equation}
\begin{equation}
d S^{(n)} \equiv \left( exp(-\epsilon^n d i_{G_n}) - 1 \right)\Theta^{(n-1)} + d S^{(n-1)}.
\end{equation}
\end{subequations}
Each term in $\vartheta^{(n)}$ can then be expressed relative to those in the previous $\vartheta^{(n-1)}$ by the following equations,
\begin{eqnarray} 
\vartheta^{(n)}_m = \sum_{k=0}^{\left[ m/n \right]} \frac{(-1)^k}{k!} \left( i_{G_n} d \right)^k \vartheta^{(n-1)}_{m-nk}
\label{eq:vartheta_n},
\end{eqnarray}
where $ \left[m/n \right] $ denotes the largest integer smaller than or equal to $m/n$.  For convenience, we define $\Theta^{(0)} \equiv \Theta$ and $\vartheta^{(0)}_m \equiv \vartheta_m, m \ge 0$. Insights of how each term in $\Theta^{(n-1)}$ is transferred by the flow of $G_n$ can be gained from Eq.(\ref{eq:vartheta_n}) which we call the transformation equations of HLTP. It can be verified that these transformation equations, if recursively expressed in terms of the previous $\vartheta^{(n-1)}$ until we obtain a relation of $\vartheta^{(n)}$,  $\vartheta^{(0)}$ and all the $G_n$'s, generate the same expressions as in Ref.~\onlinecite{littlejohn1982hamiltonian}. Of these transformation equations for arbitrary $n$, the first observation, or rather a foregone conclusion since we intentionally select $s=-\epsilon^n$ for each group of self-diffeomorphism $\mathcal{F}^s_n$, is that the $m$th-order term in $\vartheta^{(n-1)}$ for $m<n$ remains unchanged by $G_n$, i.e. that
\begin{eqnarray} 
\vartheta^{(n)}_m = \vartheta^{(n-1)}_m = \cdots = \vartheta^{(m)}_m , m< n
\label{eq:vartheta_n_concl1}.
\end{eqnarray}
The second observation is that the $n$th-order term in $\vartheta^{(n)}$ depends on the $n$-th order vector $G_n$ linearly. Taking $m=n$ in the transformation equations Eq.(\ref{eq:vartheta_n}) and using Eq.(\ref{eq:vartheta_n_concl1}) to make the superscript number of the terms $\vartheta_{m-nk}^{(n-1)}$ as small as possible, the term $\vartheta^{(n)}_n$  for $n \ge 1$ can be written as,
\begin{eqnarray} 
\vartheta^{(n)}_n  = \vartheta^{(n-1)}_n - i_{G_n} d \vartheta_0
\label{eq:vartheta_n_concl2}.
\end{eqnarray}
Assuming inductively that $\vartheta^{(n-1)}$ is known to us up to arbitrary order, the basic strategy of HLTP transformation can be divided into three steps: (\romannumeral1) the first one is to construct the one-form $\vartheta^{(n)}_n$ satisfying the relation,
\begin{eqnarray} 
L_{\zeta} d \vartheta^{(n)}_n = 0;
\label{eq:simmetry_condi_nth}
\end{eqnarray}
(\romannumeral2) the second is to solve the $G_n$ from Eq.~(\ref{eq:vartheta_n_concl2}); and finally (\romannumeral3) $\vartheta^{(n)}$ can be calculated from Eq.~(\ref{eq:vartheta_n}) up to arbitrary order once $G_n$ is known to us. It is seen that both the construction of $\vartheta^{(n)}_n$ and the determination of $G_n$ are significantly affected by the two-form $d \vartheta_0$ which has at least one null vector as explained in Sec.~\ref{sec:Preliminaries}. Taking the component of Eq.~(\ref{eq:vartheta_n_concl2}) along the direction of its null vector, say $Y$, we obtain 
\begin{eqnarray} 
 i_Y \vartheta^{(n)}_n  =  i_Y \vartheta^{(n-1)}_n + i_{G_n} i_Y d \vartheta^{(0)}_0 =  i_Y \vartheta^{(n-1)}_n
\label{eq:vartheta_n_concl3},
\end{eqnarray}
where use has been made of the relation, $i_X i_Y = - i_Y i_X$ for arbitrary vectors $X$ and $Y$. The existence of such null direction implies that the set of the linear algebraic equations for determining every $G_n$ is a singular one and that an extra restriction along the direction of the null vector, other than the one along the symmetry vector $\zeta$, must be imposed on how we could possibly prescribe each $\vartheta^{(n)}_n$. Note here that a two-form having as least null vectors as possible, that is, exactly one in the odd-dimensional $\mathcal{M}$, is often called non-singular. We therefore split our discussions into two cases depending on if the two-form $d \vartheta$ is non-singular.

\subsection{\label{sec:case1}Non-singular $d \vartheta_0$}
In this case, the theoretical feasibility of constructing symmetric $\vartheta^{(n)}_n$ and subsequently solving $G_n$ is easily guaranteed by induction. Apparently, Eq.(\ref{eq:vartheta_n_concl2}) is an equation of tensors, and, as is often the case when treating such equation, one is at liberty to represent it in any compatible coordinate system; the compatibility of coordinate systems on $\mathcal{M}$ naturally ensures that, if Eq.(\ref{eq:vartheta_n_concl2}) is valid in one of the coordinate sets, it is valid in any of them. The only restriction on the choice of coordinates is perhaps that they should not depend on the generating vector $G_n$ itself, so as not to change the linear nature of Eq.(\ref{eq:vartheta_n_concl2}). The coordinate system $\{u^{\alpha} \}$ in which $\Theta$ is originally known to us is preferred as a starting point. Therefore, the following discussion is made under the inductive assumption that $\Theta^{(n-1)}$ is known to us up to arbitrary order in coordinate system $\{u^{\alpha} \}$.

Let $Y_0$ be a non-vanishing null vector of $d \vartheta_0$. Then $Y_0$ must have non-vanishing components along at least one of the vector basis $\{\partial / \partial u^{\alpha} \}$]; assuming $\langle Y_0, du^0 \rangle \ne 0$, $Y_0$ together with $\{ Y_{\alpha} \equiv \partial / \partial u^{\alpha},1 \le \alpha \le 6 \}$ forms a new vector basis of $T \mathcal{M}$; a new dual basis $\{ \eta^{\alpha},0 \le \alpha \le 6 \}$ of $T^* \mathcal{M}$ can then be determined by solving $\langle Y_{\alpha}, \eta^{\beta} \rangle= \delta_{\alpha}^{\beta}$. The advantages of the dual basis $\{Y_{\alpha}\}$ and $\{\eta^{\alpha}\}$ are easily seen; in such dual basis we can express  $d \vartheta_0$ as 
\begin{eqnarray} 
d \vartheta_0 = \frac{1}{2} \Omega_{\alpha \beta} \eta^{\alpha} \wedge \eta^{\beta},
\label{eq:dtheta0_components}
\end{eqnarray}
where $\Omega_{\alpha \beta}$ is a $7 \times 7$ anti-symmetric matrix of functions obtained self-consistently by the relation $\Omega_{\alpha \beta} \equiv i_{Y_{\beta}} i_{Y_{\alpha}} d \vartheta_0$, and therefore must satisfy the following two properties, (\romannumeral1) $\Omega_{0 \alpha} = \Omega_{\alpha 0} = 0$, for $0 \le \alpha \le 6$ and (\romannumeral2) $\Omega_{\alpha \beta}$ for $1 \le \alpha, \beta \le 6$ is a $6 \times 6$ non-singular matrix. Consequently, the components of Eq.(\ref{eq:vartheta_n_concl2}) are divided into two parts, one serving as a constraint,
\begin{eqnarray} 
0 = i_{Y_0} \left( \vartheta_n^{(n)} - \vartheta_n^{(n-1)} \right),
\label{eq:case1_constr}
\end{eqnarray}
and the other directly solvable by inverting a matrix,
\begin{eqnarray} 
\Omega_{\alpha \beta}G_{n}^{\beta} = i_{Y_{\alpha}} \left( \vartheta_n^{(n)} - \vartheta_n^{(n-1)} \right),
\label{eq:case1_solvable}
\end{eqnarray}
where $G_{n}^{\beta} \equiv \langle G_{n}, \eta^{\beta} \rangle$. The constraint Eq. (\ref{eq:case1_constr}) somewhat conflicts with the requirement that $\vartheta^{(n)}_n$ is symmetric along $\zeta$; a necessary condition for the latter is that $i_{Y_0} \vartheta^{(n)}_n$ is also symmetric along $\zeta$, whereas the $i_{Y_0} \vartheta^{(n-1)}_n$, known to us only by inductive assumption, is generally not. To settle such conflict, an exact one-form accounting for the asymmetric in $i_{Y_0} \vartheta^{(n-1)}_n$ must be included, in addition to an one-form invariant over $\zeta$, into the definition of $\vartheta^{(n)}_n$. That is to say, the admissible choices of $\vartheta^{(n)}_n$ are restricted to the form,
\begin{eqnarray} 
\vartheta^{(n)}_n = \bar{\vartheta}_n + d \tilde{S}_n.
\label{eq:varnn_construct}
\end{eqnarray}
The $\tilde{S}_n$, a scalar function on $\mathcal{M}$, is the solution to the equation,
\begin{eqnarray} 
i_{Y_0} d \tilde{S}_n = i_{Y_0} \vartheta^{(n-1)}_n - \langle i_{Y_0} \vartheta^{(n-1)}_n \rangle,
\label{eq:case1_gauge_term}
\end{eqnarray}
where the bracket $\langle \cdot \rangle$ of a scalar function stands for its average over one cyclic period of the angle-like variable $\xi$ and should not be confused with the bilinear bracket $\langle , \rangle$ of tensors. In accordance with the constraint Eq. (\ref{eq:case1_constr}), the component of $\bar{\vartheta}_n$ along $Y_0$ must be set equal to the average of $i_{Y_0} \vartheta^{(n-1)}_n$,
\begin{eqnarray} 
i_{Y_0} \bar{\vartheta}_n = \langle i_{Y_0} \vartheta^{(n-1)}_n \rangle,
\end{eqnarray}
while the other components $i_{Y_{\alpha}} \bar{\vartheta}_n$, $1 \le \alpha \le 6$, can be in principle chosen as any scalar functions independent of $\xi$ without affecting the solvability of Eq.(\ref{eq:case1_solvable}), which is now rewritten as
\begin{eqnarray} 
\Omega_{\alpha \beta}G_{n}^{\beta} = i_{Y_{\alpha}} \left( \bar{\vartheta}_n + d \tilde{S}_n- \vartheta_n^{(n-1)} \right).
\label{eq:case1_Gn_solv}
\end{eqnarray}
It's observed that $G_n^0 \equiv \langle G_n, \eta^0 \rangle$ does not appear in any of the above equations. As a result, $G_n^0$ can be chosen as any scalar functions independent on $\xi$, which justifies the convention of setting $\langle G_n, dt \rangle = 0$ in most applications of Lie-transformation perturbation method. Now that $G_n$ is determined, $\vartheta^{(n)}$ up to arbitrary order is known to us by Eq.(\ref{eq:vartheta_n}), which by induction guarantees the solvability of all the $G_n$'s, for any integer $n \ge 1$, if all the $\vartheta^{(n)}_n$'s are constructed as shown above.

However, we must emphasize that the practicality of the above procedure is not guaranteed for the most general non-singular $d \vartheta_0$, because, as the readers may have already noticed, Eq.(\ref{eq:case1_gauge_term}) could be a 7-dimensional partial differential equations whose explicit solutions cannot be practically found. The formidability of Eq.(\ref{eq:case1_gauge_term}) characterizes those perturbation systems whose unperturbed counterpart is in its own right a complicated one and thus has a unperturbed vortex vector $Y_0$ that is not directly integrable. A typical example is the problem of charged particle motion in strong magnetic field that is subjected both to the weak non-uniformity in configuration space, often indicated by small parameters $\epsilon_B$, and to the time-dependent low-amplitude electromagnetic fluctuations, indicated by $\epsilon_{\delta}$. Through a preparatory transformation\cite{littlejohn1983variational}, the Poincar\'e-Cartan one-form governing the perturbation motion of charged particle can be cast into the form of Eq.(\ref{eq:DF_Theta}) with $\epsilon = \epsilon_{\delta}$ and $\theta_0$ in a power series of $\epsilon_B$. Accordingly, the vortex vector $Y_0$ is also in a power series of $\epsilon_B$ whose leading term is aligned with the direction of gyro-motion, and therefore the solution to Eq.(\ref{eq:case1_gauge_term}) can be approximated from a direct integration of the right hand side over the gyro-angle. But the consequence of the error within is seldom studied. Suppose the $\tilde{S}_1$ which by the previous assumption is of order $o(\epsilon^0)$ is solved approximately from Eq.(\ref{eq:case1_gauge_term}) with an error of order $o(\epsilon)$. The generating vector $G_1$ calculated from Eq.(\ref{eq:case1_Gn_solv}) will also bear an error of order $o(\epsilon)$, and hence a term of $o(\epsilon^2)$ will be unknown to us in $\Theta^{(1)} = exp(-\epsilon L_{G_1})\Theta$, making the calculations of all the succeeding $G_n$'s problematic (though it seems true that this problem can be automatically resolved for and only for $n=2$ by a re-definition of $G_2$, since $G_1$ enter linearly the determining equation for $G_2$). We make no attempt to go into details on this subject, for we believe the perturbation system of charged particle naturally fit into the singular case as will be discussed below, and therein the problem of non-integrable null vector can be circumvented for this particular perturbation system (but unfortunately not for the general ones).

\subsection{\label{sec:case2}Singular $d \vartheta_0$}
In this case, the total $d \Theta$ is also singular when $\epsilon=0$ but not so whenever $\epsilon \neq 0$, and for such $\Theta$ it can be shown that the admissible choice of $\theta^{(n)}_n$ with condition Eq.~(\ref{eq:simmetry_condi_nth}) does not always exist. Exception can be constructed where the total $\Theta$ satisfies all the previous assumption but cannot be mapped into a $\Theta^{(1)}$ that possesses the symmetry in order of $o(\epsilon)$, though it is hard to tell whether such $\Theta$ of mathematical construction corresponds to any perturbation system of physical concern. Some extra conditions must be assumed for the initial $\Theta$ so as to rule out these exceptions. To elaborate on this point, it proves necessary to discuss at first the properties of the null vectors of $d \vartheta_0$.

Suppose the non-vanishing vectors $ Y_0, Y_1, \cdots, Y_{\alpha_0} $, for some integer $\alpha_0$, $1 \le \alpha_0 \le 6$, are linearly independent null vectors of $d \vartheta_0$. The first property of these null vectors is that the maximal $\alpha_0$ is even. To prove this, we can find some other vectors among the vector basis $\{ \partial / \partial u^{\alpha} \}$, denoted by $\{Y_{\alpha_0+1}, \cdots, Y_6\}$, which are linearly independent on $Y_{\alpha}$, $\alpha \le \alpha_0$, to establish a new vector basis of $T \mathcal{M}$, and hence a dual basis $\eta^{\beta}$ of $T^* \mathcal{M}$ satisfying $\langle Y_{\alpha}, \eta^{\beta} \rangle= \delta_{\alpha}^{\beta}$. In such dual basis, the two-form $d \vartheta_0$ can be represented in similar form as in Eq.(\ref{eq:dtheta0_components}), but with $\Omega_{\alpha \beta} = \Omega_{\beta \alpha} = 0$ for $ \alpha \le \alpha_0$ and $0 \le \beta \le 6$. Assuming $\alpha_0$ is odd, the bottom right ($\alpha, \beta > \alpha_0$) part of $\Omega_{\alpha \beta}$ is an odd-dimensional anti-symmetric matrix which must have at least one eigenvector of zero eigenvalue and thus provide us with the new null vectors independent on $Y_\alpha, \alpha \le \alpha_0$. If indeed we obtain exactly one more null vectors, the proof is completed; otherwise, we may repeat the procedure above until we find either that the bottom right part of $\Omega_{\alpha \beta}$ is an even-dimensional non-singular matrix or that $d \vartheta_0 = 0$, both of these cases indicating that the maximal $\alpha_0$ is even. Let $\alpha_0$ be the maximal and we do not consider the trivial case $\alpha_0=6$. Then, all the null vectors of $d \vartheta_0$ is a linear combinations of $\{Y_0, \cdots Y_{\alpha_0} \}$, forming a $(\alpha_0 + 1)$-dimensional subspace $\mathcal{N} \subset T \mathcal{M}$. We define $\mathcal{N}_{\perp}$ as the subspace of $T \mathcal{M}$ spanned by the remaining vectors $\{Y_{\alpha_0+1}, \cdots, Y_6 \}$, and subsequently the dual subspaces $\mathcal{N}^*$ and $\mathcal{N}_{\perp}^*$ in $T^* \mathcal{M}$ can be defined as spanned respectively by the one-forms $\{ \eta_0, \cdots, \eta_{\alpha_0} \}$ and $\{ \eta_{\alpha_0+1}, \cdots, \eta_{6} \}$. The subspace $\mathcal{N}$ is called the null space (or kernel) of $d \vartheta_0$ because the two-form $d \vartheta_0$, as any two-form on $\mathcal{M}$, determines a linear transformation from $T \mathcal{M}$ to $T^* \mathcal{M}$ by the relation,
\begin{eqnarray} 
\Delta \vartheta = i_G d \vartheta_0,
\label{eq:nth_difference}
\end{eqnarray}
where $G \in T \mathcal{M}$ is any generating vector, and the one-form $\Delta \vartheta \in T^* \mathcal{M}$, according to Eq.(\ref{eq:vartheta_n_concl2}), measures the effective difference of $n$th-order term between $\Theta^{(n-1)}$ and $\Theta^{(n)}$. Owing to the singularity of $d \vartheta_0$, the input $G$ in $\mathcal{N}$ is not involved in such linear transformation, and the output $\Delta \vartheta$ must fully belong to $\mathcal{N}^*_{\perp}$, implying that $\vartheta^{(n)}_{n}$ is identical to $\vartheta^{(n-1)}_{n}$ when both projected into $\mathcal{N}^*$. In consequence, if any asymmetry along $\zeta$ exists in the components of $\vartheta^{(n-1)}_{n}$ in $\mathcal{N}^*$, it must be preserved instead of being eliminated in $\vartheta^{(n)}_{n}$.  Setting $\alpha_0=0$, we return to the previous non-singular case. In such case, the asymmetry of $\vartheta^{(n-1)}_{n}$ is not eliminated either, but can be absorbed in one single inconsequential term $d \tilde{S_n}$, for the dimension of $\mathcal{N}^*$ is exactly one. This approach can not be directly exploited when $d \vartheta_0$ has multiple null vectors. Therefore, to obtain the very first generating vector $G_1$ and initiate the HLTP transformation, one must assume that the first-order term $\vartheta_1$ in $\Theta$ is symmetric in the subspace $\mathcal{N}^*$ (to be defined rigorously in Eq.(\ref{eq:extra_assump1}) ). Once such assumption is made on $\vartheta_1$, the $G_1$ in $\mathcal{N}_{\perp}$ is completely determined by projecting Eq.(\ref{eq:vartheta_n_concl2}) into $\mathcal{N}^*_{\perp}$, for the two-form $d \vartheta_0$ by Eq.(\ref{eq:nth_difference}) establishes a homomorphism between $\mathcal{N}_{\perp}$ and $\mathcal{N}^*_{\perp}$. The question then is how to construct all the succeeding $\vartheta^{(n)}_n$'s and guarantee the solvability of $G_n$'s. A clue can be obtained considering the fact that $d \vartheta_0$ determines a degenerate linear transformation from $\mathcal{N}_{\perp}$ to $\mathcal{N}^*_{\perp}$ rather than from $T \mathcal{M}$ to $T^* \mathcal{M}$. Therefore, we must find another two-form to compensate for the singularity of $d \vartheta_0$ so as to map the generating vector in $\mathcal{N}$ into an one-form in $\mathcal{N}^*$. Another property of the subspace $\mathcal{N}$ is needed to clarify this point.

The second property of the null vectors of $d \vartheta_0$ is that the Poisson bracket of any two vectors in $\mathcal{N}$ is still a vector in $\mathcal{N}$. This is easily proved with the help of the Cartan formulas Eq. (\ref{eq:cartan_formulas_1}) and (\ref{eq:cartan_formulas_2}). Since
\begin{eqnarray} 
i_{[Y,Z]} d \vartheta_0 &=& L_Y i_Z d \vartheta_0 - i_Z L_Y d \vartheta_0 \nonumber \\
&=& L_Y i_Z d \vartheta_0 - i_Z d i_Y d  \vartheta_0,
\end{eqnarray}
the vector $[Y,Z]$ is a null vector of $ d \vartheta_0$ whenever $Y$ and $ Z$ are both null vectors of $ d \vartheta_0$. Therefore, the Frobenius theorem (see Ref.~\onlinecite{chern1999lectures}) applies to the null vectors $\{ Y_0, \cdots, Y_{\alpha_0} \}$ of $d \vartheta_0$, ensuring the existence of such a coordinate set $\{ w^{\alpha}, 0 \le \alpha \le 6 \}$ that $\mathcal{N}$ and $\mathcal{N}_{\perp}^*$ are spanned respectively by $\{\partial / \partial w^0, \cdots, \partial / \partial w^{\alpha_0} \}$ and $\{ d w^{\alpha_0+1}, \cdots, d w^{6}\}$. Although in general the coordinate set $\{ w^{\alpha} \}$ is not obtained without performing some non-trivial integrations, the mere knowledge of its existence will suffice and we can assume there is a $(\alpha_0 + 1) \times (\alpha_0 + 1)$ non-singular matrix $R_{\alpha}^{\beta}$ such that, for $0 \le \alpha , \beta \le \alpha_0$,
 \begin{eqnarray} 
Y_{\alpha} = R_{\alpha}^{\beta} \frac{\partial}{\partial w^{\beta}}.
\label{eq:frobenius_th}
\end{eqnarray}
In addition, our discussions below profit a lot if all the vector basis $\{ Y_{\alpha} \}$ remain to be invariant of $\zeta$. Provided that the component matrix of $d \vartheta_0$ under the original basis $\{ d u^{\alpha} \wedge d u^{\beta} \}$ is composed of functions independent on $\xi$, the null vectors of $d \vartheta_0$ we can directly obtain from such matrix must have components independent on $\xi$ in the original basis $\{ \partial / \partial u^{\alpha} \}$. Therefore, without loss of generality we can assume that, for $0 \le \alpha \le \alpha_0$,
 \begin{eqnarray} 
[Y_{\alpha}, \zeta] = 0,
\label{eq:commute_zetay}
\end{eqnarray}
which guarantee the commutativity between $L_{\zeta}$ and all the null vectors according to Eq.(\ref{eq:cartan_formulas_2}).

It can now be verified that the linear transformation determined by $d \vartheta_1$ must be as non-degenerate as possible (to be defined rigorously in Eq.(\ref{eq:extra_assump2})), in order for the HLTP approach to be valid in the singular case. By the second property aforementioned, we can find some functions $f_{\alpha}$ on $\mathcal{M}$, for $\alpha>\alpha_0$, such that
 \begin{eqnarray} 
i_{G} d \vartheta_0 = f_{\alpha} d w^{\alpha} \in \mathcal{N}^*_{\perp}.
\end{eqnarray}
For such an one-form generated by any vector $G$, its exterior derivative, $d i_{G} d \vartheta_0$, must have a component matrix in the basis $\{ d w^{\alpha} \wedge d w^{\beta} \}$ that is empty in the upper left ($\alpha,\beta \le \alpha_0$). Considering Eq.(\ref{eq:frobenius_th}) and the linearity of the operator $i$, this implies that
 \begin{eqnarray} 
i_{Y_{\alpha}} i_{Y_{\beta}} di_{G} d \vartheta_0 = 0, \quad \alpha, \beta \le \alpha_0.
\label{eq:fb_concl}
\end{eqnarray}
Moreover, taking the exterior derivative of both side in Eq.(\ref{eq:vartheta_n_concl2}), it is seen that the upper left part of the component matrix of $d \vartheta^{(n)}_{n}$ in the basis $\{ \eta^{\alpha} \wedge \eta^{\beta} \}$ is identical to that of $d \vartheta^{(n-1)}_{n}$ in the same basis, i.e.
\begin{eqnarray} 
i_{Y_{\alpha}} i_{Y_{\beta}} d \vartheta^{(n)}_{n} = i_{Y_{\alpha}} i_{Y_{\beta}} d \vartheta^{(n-1)}_{n}, \quad \alpha, \beta \le \alpha_0. 
\label{eq:exception1}
\end{eqnarray}
The implication of Eq.(\ref{eq:exception1}) is twofold. One is that it demonstrates again the asymmetry of $\vartheta^{(n-1)}_{n}$ is preserved in $\vartheta^{(n)}_{n}$ in the subspace $\mathcal{N}^*$, for, taking the Lie derivative of both side in Eq.(\ref{eq:exception1}) and considering the commutativity Eq.(\ref{eq:commute_zetay}), it is seen that the two form $L_{\zeta} d \vartheta^{(n)}_n$ is identical to $L_{\zeta} d \vartheta^{(n-1)}_n$ when both projected into the subspace spanned by the two-forms $\{ \eta^{\alpha} \wedge \eta^{\beta}; \alpha, \beta \le \alpha_0\}$. The other implication is seen when $\vartheta^{(n+1)}_{n+1}$ is expressed into terms of $\vartheta^{(n-1)}$. According to the transformation equations Eq.(\ref{eq:vartheta_n}), we have, for $n \ge 2$,
\begin{eqnarray} 
\vartheta^{(n+1)}_{n+1} &=& \vartheta^{(n)}_{n+1} - i_{G_{n+1}} d \vartheta_0 \nonumber \\
&=& \vartheta^{(n-1)}_{n+1} - i_{G_{n}} d \vartheta^{(1)}_1 -  i_{G_{n+1}} d \vartheta_0.
\label{eq:vtn1_nlg2}
\end{eqnarray}
Projecting the equation above into the subspace $\mathcal{N}^*$, it is seen that $G_n$ in $\mathcal{N}$ which is left unused in constructing $\vartheta^{(n)}_{n}$ in $\mathcal{N}^{*}_{\perp}$ can be exploited in constructing the next-order $\vartheta^{(n+1)}_{n+1}$ in $\mathcal{N}^*$. The two-form $d \vartheta^{(1)}_1$ is then required to determine a linear transformation from $\mathcal{N}$ to $\mathcal{N}^*$ complementing the one determined by $d \vartheta_0$ from $\mathcal{N}_{\perp}$ to $\mathcal{N}^*_{\perp}$. According to Eq.(\ref{eq:exception1}), whether $d \vartheta^{(1)}_1$ satisfies such requirement depends solely on $d \vartheta_1$.

To sum up the above discussions, the feasibility of HLTP approach in the singular case relies on two extra conditions for the first-order term $\vartheta_1$ which are listed as follows:
\begin{itemize}
\item[(\romannumeral1)] the $\vartheta_1$ in the subspace $\mathcal{N}^*$ is symmetric along $\zeta$, implying the existence of a scalar function $\tilde{S}_1$ such that  
\begin{eqnarray} 
L_{\zeta} (\vartheta_1 - d \tilde{S}_1) \in \mathcal{N}^*_{\perp};
\label{eq:extra_assump1}
\end{eqnarray}
\item[(\romannumeral2)] the rank of the sub-matrix $\Lambda_{\alpha \beta} \equiv i_{Y_{\beta}} i_{Y_{\alpha}} d \vartheta_1$, for $0 \le \alpha, \beta \le \alpha_0$, is exactly $\alpha_0$, i.e.,
\begin{eqnarray} 
r \left( \Lambda \right) = \alpha_0.
\label{eq:extra_assump2}
\end{eqnarray}
\end{itemize}
If the condition (\romannumeral1) above is satisfied, the scalar function $\tilde{S}_1$ is obtained by solving the equation,
\begin{eqnarray} 
i_{Y_{\alpha}} d \tilde{S}_1 = i_{Y_{\alpha}} \vartheta_1 - \langle i_{Y_{\alpha}} \vartheta_1\rangle,
\end{eqnarray}
for one particular $\alpha \le \alpha_0$, and the $\tilde{S}_1$ thus obtained must automatically satisfy the same equation for arbitrary $\alpha \le \alpha_0$. If the condition (\romannumeral2) is satisfied, we can perform a linear transformation on the null vectors $\{Y_0, \cdots, Y_{\alpha} \}$ so that the new $\Lambda'_{\alpha \beta}$ satisfies $\Lambda'_{0 \alpha} = \Lambda'_{\alpha 0} = 0$ for $0 \le \alpha \le \alpha_0$ and the remaining part is an $\alpha_0 \times \alpha_0$ non-singular sub-matrix. For convenience, we assume such transformation has already been performed so that the $\Lambda$ satisfies the above conditions for $\Lambda'$.

In what follows the solvability of $G_n$ for every $n \ge 1$ is shown based on the inductive conditions that $\vartheta^{(n-1)}$ is known to us up to arbitrary order and that we have obtained the scalar function $\tilde{S}_n$ satisfying
\begin{eqnarray} 
L_{\zeta} (\vartheta^{(n-1)}_n - d \tilde{S}_n) \in \mathcal{N}^*_{\perp}.
\label{eq:extra_assump1}
\end{eqnarray}
The scheme of determining $G_n$ is a staggered one, for the components of $G_n$ in the subspaces $\mathcal{N}_{\perp}$ and $\mathcal{N}$ are determined, respectively, by constructing $\vartheta^{(n)}_n$ in $\mathcal{N}^*_{\perp}$ and $\vartheta^{(n+1)}_{n+1}$ in $\mathcal{N}^*$. Note that the second equals sign of Eq.(\ref{eq:vtn1_nlg2}) is valid only for $n \ge 2$, so the determining equations of $G_1$ is formally different from those of $G_n$, $n \ge 2$.  For $n=1$, we can express $\vartheta^{(2)}_2$ in terms of $G_1$ according to the transformation equations Eq.(\ref{eq:vartheta_n}),
\begin{eqnarray} 
\vartheta^{(2)}_2 = \vartheta_2 - i_{G_1} d ( \vartheta_1 - \frac{1}{2} i_{G_1} d \vartheta_0 )   - i_{G_2} d \vartheta_0.
\label{eq:vt2_nlg2}
\end{eqnarray}
Projecting Eq.~(\ref{eq:vartheta_n_concl2}) for $n=1$ into $\mathcal{N}^*_{\perp}$ and Eq.~(\ref{eq:vt2_nlg2}) into $\mathcal{N}^*$, we obtain the determining equations for $G_1$ respectively in $\mathcal{N}_{\perp}$ and $\mathcal{N}$ as in the following:
\begin{subequations}
\label{eq:sigl_G1_det}
\begin{equation}
\sum_{\beta=\alpha_0+1}^{6}  \Omega_{\alpha \beta} G_1^{\beta} = i_{Y_{\alpha}} \left( \vartheta^{(1)}_1 -  \vartheta_1 \right), \quad \alpha \ge \alpha_0 +1, \label{subeq:sigl_G1_det_1}
\end{equation}
\begin{eqnarray}
&&\sum_{\beta=0}^{\alpha_0} \Lambda_{\alpha \beta} G_1^{\beta}  = i_{Y_{\alpha}} \left( \vartheta^{(2)}_2 - \vartheta_2    \right) - \sum_{\beta=\alpha_0+1}^6 G^{\beta}_1 \Lambda_{\alpha \beta} \nonumber \\
&&  + \frac{1}{2} \sum_{\beta, \gamma=\alpha_0+1}^6 G^{\beta}_1 i_{Y_{\beta}} i_{Y_{\alpha}}d \left( G^{\gamma}_1 i_{Y_{\gamma}} d \vartheta_0 \right), \quad \alpha \le \alpha_0 .
\label{subeq:sigl_G1_det_2}
\end{eqnarray}
\end{subequations}
Note that Eq.~(\ref{eq:fb_concl}) must be used to obtain Eq.~(\ref{subeq:sigl_G1_det_2}). Similarly as in Eq.(\ref{eq:varnn_construct}) in the non-singular case, both $\vartheta^{(1)}_{1}$ and $\vartheta^{(2)}_2$ should be constructed as the addition of the symmetric term $\bar{\vartheta}_n$ and the exact term $d \tilde{S}_n$. The $\tilde{S}_1$ is obtained by the above assumption (\romannumeral1), while the $\tilde{S}_2$ is obtained from the oscillating part of Eq.(\ref{subeq:sigl_G1_det_2}) for $\alpha = 0$, i.e. from the following equation,
\begin{eqnarray}
i_{Y_0} d \tilde{S}_{2} &=&\tilde{\mathcal{G}} i_{Y_0}\left( \vartheta_{2} - \sum_{\beta=\alpha_0+1}^6 G^{\beta}_1 i_{Y_{\beta}} d \vartheta_1 \right. \nonumber \\
&&\left.  + \frac{1}{2} \sum_{\beta, \gamma=\alpha_0+1}^6 G^{\beta}_1 i_{Y_{\beta}} d \left( G^{\gamma}_1 i_{Y_{\gamma}} d \vartheta_0 \right) \right),
\label{eq:oscillating_s1}
\end{eqnarray}
where $\tilde{\mathcal{G}}$ is an operator such that $\tilde{\mathcal{G}} f \equiv f - \langle f \rangle$ for arbitrary function $f$ on $\mathcal{M}$. For $n \ge 2$, likewise, the determining equations for $G_n$ can be obtained as in the following:
\begin{subequations}
\label{eq:sigl_Gn_det}
\begin{equation}
\sum_{\beta=\alpha_0+1}^{6}  \Omega_{\alpha \beta} G_n^{\beta} = i_{Y_{\alpha}} \left( \vartheta^{(n)}_n -  \vartheta^{(n-1)}_n \right), \quad \alpha \ge \alpha_0 +1, \label{subeq:sigl_Gn_det_1}
\end{equation}
\begin{eqnarray}
\sum_{\beta=0}^{\alpha_0} \Lambda_{\alpha \beta} G_n^{\beta}  &=& i_{Y_{\alpha}} \left( \vartheta^{(n+1)}_{n+1} - \vartheta^{(n-1)}_{n+1}    \right) \nonumber \\
 &&- \sum_{\beta=\alpha_0+1}^6 G^{\beta}_n i_{Y_{\beta}} i_{Y_{\alpha}} d \vartheta^{(1)}_1, \quad \alpha \le \alpha_0.
\label{subeq:sigl_Gn_det_2}
\end{eqnarray}
\end{subequations}
Note again that we have used Eq.~(\ref{eq:exception1}) to obtain Eq.~(\ref{subeq:sigl_Gn_det_2}). The scalar function $\tilde{S}_{n+1}$ which provides the inductive condition for the next order is obtained from the oscillating part of Eq.(\ref{subeq:sigl_Gn_det_2}) for $\alpha = 0$,
\begin{eqnarray}
i_{Y_0} d \tilde{S}_{n+1} =\tilde{\mathcal{G}} i_{Y_0}\left( \vartheta^{(n-1)}_{n+1} - \sum_{\beta=\alpha_0+1}^6 G^{\beta}_n i_{Y_{\beta}}d \vartheta^{(1)}_1 \right).
\label{eq:oscillating_sn}
\end{eqnarray}
It is seen that to carry out HLTP transformation in the case of singular $d \vartheta_0$, we have to find the gauge functions $\tilde{S}_n$ as in the non-singular case. Therefore, solving the partial differential equations along one of the null vectors of $d \vartheta_0$ is generally inevitable. But we will see in the following that guiding-center motion is a special case where $Y_0 = \zeta$ and hence $\tilde{S}_n$ is obtained by direct integration.

\section{\label{sec:gc_trans} 
guiding-center motion as the singular perturbation motion}
Charged particle motion in the magnetic field is govern by the well-known PCOF which in the Cartesian coordinates $\{t, \bold{x}, \bold{v} \}$ is represented as
\begin{eqnarray}
\Theta = e \bold{A} \cdot d\bold{x} + m \bold{v} \cdot d\bold{x} - (\frac{1}{2} m \bold{v} \cdot \bold{v}  + e \phi) dt
\label{eq:dimensional_car}.
\end{eqnarray}
It can be checked directly that vortex vector $X$ of $\Theta$ is 
\begin{eqnarray}
X =\frac{\partial}{\partial t} +  \bold{v} \cdot \frac{\partial}{\partial \bold{x}} + \frac{e}{m} \left( \bold{E} + \bold{v} \times \bold{B} \right) \cdot \frac{\partial}{\partial \bold{v}}
\label{eq:dimn_car_vortex}.
\end{eqnarray}
Here the notation of quantities is conventional, and we use the dot product without involving any geometric meaning but as a simple notation of the product of $1 \times 3$ matrices which are denoted by bold letters and consist of either 3 scalar functions on $\mathcal{M}$, 3 one-forms of $T^*{\mathcal{M}}$ or 3 vectors of $T{\mathcal{M}}$; for example, $\bold{A} = (A_1, A_2, A_3)$, $d\bold{x} = (dx^1, dx^2, dx^3)$ and $\bold{A} \cdot d\bold{x} = A_1 dx^1+A_2 dx^2+A_3 dx^3$ . Although some of these matrices are indeed representations of the vectors of $T \mathcal{C}$, we insist on calling them $1 \times 3$ matrices to avoid confusion with the vectors of $T \mathcal{M}$. The cross product between these matrices can be easily defined with the help of Levi-Civita symbol. 

To introduce the small parameter $\epsilon$, we split the potentials into an equilibrium part denoted by the subscript '0', and a perturbed part denoted by '1'; for example, $\bold{A}=\bold{A_0} + \bold{A_1}$ and $\phi=\phi_0 + \phi_1$. And we perform normalizations over the coordinates, 
\begin{eqnarray}
\hat{\bold{x}} \equiv \frac{\bold{x}}{L_{ref}},\quad \hat{\bold{v}} \equiv \frac{\bold{v}}{v_{ref}}, \quad \hat{t} \equiv \frac{t v_{ref}}{L_{ref}}
\label{eq:cor_norm},
\end{eqnarray}
where $L_{ref}$ is the characteristic length of the equilibrium field and $v_{ref}$ the particle thermal velocity. Each potential is normalized by the characteristic magnitude of the field that it generates such as in the following,
\begin{eqnarray}
\hat{\bold{A}}_{\bold{1}} \equiv \frac{\bold{A_1}}{B_{1ref}L_{ref}},~ \hat{\phi}_0 \equiv \frac{\phi_0}{E_{0ref}L_{ref}}
\label{eq:potential_norm}.
\end{eqnarray}
Divided by an overall reference unit, ${e B_{0ref} L^2_{ref}}$, the PCOF becomes
\begin{eqnarray}
\hat{\Theta}  &=  &\hat{\bold{A}}_{\bold{0}} \cdot d \hat{\bold{x}} - \epsilon_{E} \hat{\phi}_0 d \hat{t} \nonumber \\
&&+ (\epsilon_{A}  \hat{\bold{A}}_{\bold{1}}  + \epsilon \hat{\bold{v}} ) \cdot d \hat{\bold{x}} -  ( \epsilon\frac{1}{2} \hat{\bold{v}} \cdot \hat{\bold{v}} + \epsilon_{\phi} \hat{\phi}_1  )d \hat{t}
\label{eq:norm_car},
\end{eqnarray}
where the constants are defined respectively as
\begin{eqnarray}
&&\epsilon \equiv \frac{m v_{ref}}{e B_{0ref} L_{ref}}, \quad \epsilon_A \equiv \frac{B_{1ref}}{B_{0ref}}, \nonumber \\ &&\epsilon_{E} \equiv \frac{E_{0ref}}{B_{0ref}v_{ref}}, \quad \epsilon_{\phi} \equiv \frac{E_{1ref}}{B_{0ref}v_{ref}}
\label{eq:df_epsilon}.
\end{eqnarray}
We assume the following orderings,
\begin{eqnarray}
\epsilon \approx \epsilon_A \approx \epsilon_{\phi}  \ll 1 \approx \epsilon_{E}
\label{eq:ordering},
\end{eqnarray}
which are consistent with the gyrokinetic ordering\cite{brizard2007foundations} in the case of strong $E \times B$ shearing. In further we require that $\bold{E_0}$ is everywhere perpendicular to $\bold{B_0}$, otherwise with parallel $\bold{E_0}$ the charged particle may obtain a large parallel velocity and undergo significant magnetic variation during a gyro-period\cite{littlejohn1981hf}, invalidating the guiding-center approximation. For the simplicity of notations, we drop the hat above a quantities, as all the quantities henceforth are dimensionless. Setting $\epsilon=\epsilon_{A} = \epsilon_{\phi}$ and $\epsilon_{E}=1$, the PCOF then becomes
\begin{eqnarray}
\Theta &=&  \bold{A}_{\bold{0}}  \cdot d \bold{x} -  \phi_0 d t  + \epsilon \left(  \bold{A}_{\bold{1}}  \cdot d \bold{x} -    \phi_1 d t  \right)\nonumber \\
&& \quad + \epsilon \left( \bold{v} \cdot d \bold{x} - \frac{1}{2} \bold{v} \cdot \bold{v}dt \right)
\label{eq:norm_car_wohat}.
\end{eqnarray}
To introduce the gyro-angle $\xi$, we define $\bold{b} \equiv \bold{B_0}/\vert \bold{B_0} \vert$, with $\vert \bold{B_0} \vert \equiv \sqrt{\bold{B_0} \cdot \bold{B_0}}$ nowhere vanishing. Subsequently, two unit matrices $\bold{a}$ and $\bold{c}$ can be found satisfying
\begin{subequations}
\label{eq:abc_relation}
\begin{eqnarray}
&&\bold{a} \cdot \bold{b} = \bold{b} \cdot \bold{c} = \bold{c} \cdot \bold{a} = 0, \\
&&\bold{a}\cdot\bold{a} =\bold{b}\cdot\bold{b} = \bold{c}\cdot\bold{c} =1= (\bold{a} \times \bold{b} ) \cdot \bold{c}.
\end{eqnarray}
\end{subequations}
Note that both $\bold{a}$ and $\bold{c}$ are independent on variables of velocity space, and that we do not consider the subtle problems such as the gyro-gauge invariant\cite{littlejohn1988phase} and the existence\cite{burby2012gyrosymmetry} of globally smooth $\bold{a}$ and $\bold{c}$. In what follows we rely essentially on $\bold{b}$, rather than on any of $\bold{a}$ and $\bold{c}$, in order to define the gyro-vector $\zeta$ and the null space $\mathcal{N}$. The parallel and perpendicular velocity $v_{\parallel}$, $v_{\perp}$and gyro-angle $\xi$ are defined by the following two relations,
\begin{subequations}
\begin{eqnarray}
v_{\parallel} &\equiv& \bold{v} \cdot \bold{b}, \label{eq:df_vpar}\\
 v_{\perp} (-cos\xi \bold{a} - sin\xi \bold{c}) &\equiv& \bold{v} -\bold{u_E}-  (\bold{v} \cdot \bold{b}) \bold{b} \label{eq:df_vperp_xi},
\end{eqnarray}
\label{eq:df_vpar_vper_xi}
\end{subequations}
where $\bold{u_E} \equiv \bold{E_0} \times \bold{b}/ \vert \bold{B_0} \vert$.  For simplicity we assume both $\bold{B_0}$ and $\bold{E_0}$ are independent on $t$, so that $\bold{a}$, $\bold{b}$ and $\bold{c}$ are all independent on $t$. Our results will be formulated in the coordinates $\{ t, \bold{x}, v_{\parallel}, v_{\perp}, \xi \}$; their exterior derivatives (which can be easily worked out from Eqs.~(\ref{eq:df_vpar_vper_xi})) form a basis of $T^* \mathcal{M}$, while correspondingly the dual basis of $T \mathcal{M}$ are given in the following,
\begin{subequations}
\label{eq:vb_trans}

\begin{flalign}
&~ \left( \frac{\partial}{\partial \bold{x}} \right)' = \frac{\partial}{\partial \bold{x}}+ \frac{\partial \bold{u_E}}{\partial \bold{x}} \cdot \frac{\partial}{\partial \bold{v}} \nonumber &\\
&\quad \quad\quad \quad\quad+  (\bold{v} - \bold{u_E}) \cdot (\bold{a}\frac{\partial \bold{a}}{\partial \bold{x}}  +  \bold{b} \frac{\partial \bold{b}}{\partial \bold{x}} +\bold{c} \frac{\partial \bold{c}}{\partial \bold{x}} )\cdot \frac{\partial}{\partial \bold{v}} ,&
\label{subeq:px_prime}\\
&\quad\quad \frac{\partial}{\partial v_{\parallel}}= \bold{b} \cdot \frac{\partial}{\partial \bold{v}},& \\
&\quad\quad \frac{\partial}{\partial v_{\perp}} = \frac{\bold{v} - \bold{u_E}- \bold{v} \cdot \bold{b} \bold{b}}{v_{\perp}} \cdot \frac{\partial}{\partial \bold{v}}, &\\
&\quad\quad ~~ \frac{\partial}{\partial \xi} = \left( (\bold{v}- \bold{u_E}) \times \bold{b} \right) \cdot \frac{\partial}{\partial \bold{v}}.&
\end{flalign}
\end{subequations}
The complexity of Eq.~(\ref{subeq:px_prime}) won't cause any trouble, for $(\partial / \partial \bold{x})'$ acting on the potentials is effectively identical to $\partial / \partial \bold{x} \equiv \nabla$. Therefore, when formulating our results we do not tell the difference between $(\partial / \partial \bold{x})'$ and $\partial / \partial \bold{x}$, and one should bear in mind that the $\partial / \partial \bold{x}$ hereafter denotes the natural vector basis of the new coordinates $\{ t, \bold{x}, v_{\parallel}, v_{\perp}, \xi \}$ rather than that of the original $\{ t, \bold{x}, \bold{v} \}$. The PCOF can be cast into the form of $\Theta = \vartheta_0 + \epsilon \vartheta_1$ with $\vartheta_0$ and $\vartheta_1$ defined respectively as
\begin{subequations}
\label{eq:norm_xi_eps}
\begin{eqnarray}
 &&\vartheta_0 \equiv  \bold{A}_{\bold{0}}  \cdot d \bold{x} -  \phi_0 d t, \\
&&\vartheta_1 \equiv   \left(  \bold{A}_{\bold{1}}  + \bold{u_E} +  v_{\parallel} \bold{b} + v_{\perp} \bold{\tilde{c}} \right)  \cdot d \bold{x} \nonumber \\
&& \quad - \frac{1}{2} \left( v_{\parallel}^2+v_{\perp}^2 +2 v_{\perp} \bold{u_E} \cdot \bold{\tilde{c}}  + \bold{u_E}\cdot \bold{u_E}+ \phi_1\right)dt .
\end{eqnarray}
\end{subequations}
The vortex vector $X$ in the natural basis of $\{ t, \bold{x}, v_{\parallel}, v_{\perp}, \xi \}$ is represented as
\begin{eqnarray}
X &=& \frac{\partial}{\partial t} + \left( \bold{u_E}+v_{\parallel} \bold{b} + v_{\perp} \bold{\tilde{c}} \right ) \cdot \frac{\partial}{\partial \bold{x}} \nonumber \\ 
&&+ \left( \bold{\tilde{E}_1}\cdot \bold{b} +v_{\perp} \bold{\tilde{B}_1}\cdot \bold{\tilde{a}}\right) \frac{\partial}{\partial v_{\parallel}} + \left(  \bold{\tilde{E}_1}  \cdot \bold{\tilde{c}} - v_{\parallel}\bold{\tilde{B}_1} \cdot \bold{\tilde{a}} \right)  \frac{\partial}{\partial v_{\perp}} \nonumber \\
   && + \left( \frac{1}{\epsilon} \vert \bold{B_0} \vert - \frac{1}{v_{\perp}} \bold{\tilde{E}_1} \cdot \bold{\tilde{a}} + \bold{\tilde{B}_1} \cdot ( \bold{b} - \frac{v_{\parallel}}{v_{\perp}} \bold{\tilde{c}}) \right)  \frac{\partial}{\partial \xi}
\label{eq:vortex_diml_vpxi},
\end{eqnarray}
where uses have been made of the following definitions,
\begin{subequations}
\begin{eqnarray}
\bold{\tilde{a}} &\equiv& - sin \xi \bold{a} + cos \xi \bold{c}, \\
\bold{\tilde{c}} &\equiv& - cos \xi \bold{a} - sin \xi \bold{c} , \\
\bold{\tilde{B}_1} & \equiv &   \bold{B_1} +  v_{\parallel} \nabla \times \bold{b} + v_{\perp} \nabla \times \bold{\tilde{c}} + \nabla \times \bold{u_E}, \\
\bold{\tilde{E}_1} &\equiv& \bold{E_1} - \nabla \left( v_{\perp}\bold{u_E} \cdot \bold{\tilde{c}}  + \frac{1}{2} \bold{u_E}\cdot \bold{u_E}\right) + \bold{u_E} \times \bold{\tilde{B}_1}.
\end{eqnarray}
\end{subequations}
We remark that the vector Eq.(\ref{eq:vortex_diml_vpxi}) can be compared with Eq.(\ref{eq:dimn_car_vortex}), for they are two different representations of the same vortex vector $X$ considering the normalization and the split of the potentials.

We can therefore apply what has been discussed in Sec.~\ref{sec:general} to the PCOF defined in Eq.~(\ref{eq:norm_xi_eps}). To begin with, it's easy to tell that the $d \vartheta_0$ has far more than one null vector (because the one-form $\vartheta_0 \in T^*\mathcal{M}$ in essence is an one-form induced by the projection from $\mathcal{M}$ to $R \times \mathcal{C}$). We simply list all the linearly independent null vectors of $d \vartheta_0$:
\begin{eqnarray}
Y_{t} = \frac{\partial}{\partial t} + \bold{u_E} \cdot \frac{\partial}{\partial \bold{x}}, \quad Y_{b} = \bold{b} \cdot \frac{\partial}{\partial \bold{x}}, \nonumber \\
Y_{\parallel} = \frac{\partial}{\partial v_{\parallel}}, \quad Y_{\perp} = \frac{\partial}{\partial v_{\perp}}, \quad \zeta \equiv Y_{\xi} = \frac{\partial}{\partial \xi},
\label{eq:null_vectors}
\end{eqnarray}
The remaining two vectors that are linearly independent on those in Eqs.(\ref{eq:null_vectors}) can be selected as 
\begin{eqnarray}
Y_{a} = \bold{\tilde{a}}\cdot \frac{\partial}{\partial \bold{x}}, \quad Y_{c} = \bold{\tilde{c}}\cdot \frac{\partial}{\partial \bold{x}}
\label{eq:not_null_vectors}.
\end{eqnarray}
The null space $\mathcal{N}$ of $d \vartheta_0$ is then spanned by $\{Y_t, Y_b, Y_{\parallel}, Y_{\perp}, Y_{\xi} \}$, while another subspace $\mathcal{N}_{\perp}$ is spanned by $\{ Y_a, Y_c \}$. Other selections of $\{ Y_a, Y_c \}$ such as substituting the $\bold{\tilde{a}}$ and $\bold{\tilde{c}}$ in Eq.(\ref{eq:not_null_vectors}) with $\bold{a}$ and $\bold{c}$ respectively are possible and should not affect all the $G_n$'s to be obtained. Generally speaking, any matrices $\bold{a}$ and $\bold{c}$ that satisfy $(\bold{a} \times \bold{b}) \cdot \bold{c} \neq 0$ can be selected to define the subspace $\mathcal{N}_{\perp}$, but those in Eqs.~(\ref{eq:not_null_vectors}) prove to be most convenient. In particular, we note that both the $Y_t$ and $Y_b$ are null vectors of $d \vartheta_0$ only when we assume $\bold{E_0} \cdot \bold{B_0} = 0$. To assume otherwise not only jeopardizes the guiding-center approximation as mentioned above but ruins the global property of $\mathcal{N}$ which is $5$-dimensional at points of $\mathcal{M}$ where $\bold{E_0} \cdot \bold{B_0} = 0$ but shrinks to be $3$-dimensional where $\bold{E_0} \cdot \bold{B_0} \neq 0$. 

As the $d \vartheta_0$ is singular, we need to check if those two conditions in Sec.~\ref{sec:case2} are satisfied. Replacing the numeric indexes of $Y_{\alpha}$ in Sec.~\ref{sec:case2} with the symbols $\alpha \in \{ t, a, b, c, \parallel, \perp, \xi \}$, the dual basis $\{ \eta^{\beta} \}$ can be expressed as follows: those spanning $\mathcal{N}^*$ are
\begin{eqnarray}
&&\quad \eta^{t} = dt, \quad \eta^{b} = \bold{b} \cdot d \bold{x}, \nonumber \\
 && \eta^{\parallel} = d v_{\parallel}, \quad \eta^{\perp} = d v_{\perp},\quad \eta^{\xi} = d \xi,
\label{eq:null_vector_dual}
\end{eqnarray}
and the others spanning $\mathcal{N}^*_{\perp}$ are
\begin{eqnarray}
\eta^{a} = \bold{\tilde{a}} \cdot d \bold{x} - (\bold{u_E} \cdot \bold{\tilde{a}}) d t, \quad \eta^{c} = \bold{\tilde{c}} \cdot d \bold{x} - (\bold{u_E} \cdot \bold{\tilde{c}} )d t.
\label{eq:non_null_vector_dual}
\end{eqnarray}
By direct calculation, the Lie derivative of $\vartheta_1$ along $\zeta$ is 
\begin{eqnarray}
L_{\zeta} \vartheta_1 = - v_{\perp} \bold{\tilde{a}} \cdot d \bold{x} +  v_{\perp}\bold{u_E} \cdot \bold{\tilde{a}} dt = -v_{\perp} \eta^{a} \in \mathcal{N}^*_{\perp},
\end{eqnarray}
which means the condition (\romannumeral1) is trivially satisfied with $\tilde{S}_1=0$. The $7 \times 7$ anti-symmetric matrices $\Omega_{\alpha \beta} \equiv i_{Y_{\beta}}i_{Y_{\alpha}} d \vartheta_0$ and $\Lambda_{\alpha \beta} \equiv i_{Y_{\beta}}i_{Y_{\alpha}} d \vartheta_1$ can be easily worked out from the following expressions,
\begin{eqnarray}
d \vartheta_0 &=& \vert \bold{B_0} \vert \eta^{c} \wedge \eta^{a}, \\
d \vartheta_1 &=& \bold{\tilde{E}_1} \cdot (\bold{\tilde{a}} \eta^{a} + \bold{b} \eta^{b} + \bold{\tilde{c}} \eta^{c})\wedge \eta^t \nonumber \\
& +&\bold{\tilde{B}_1} \cdot (\bold{\tilde{a}} \eta^{b} \wedge \eta^{c} + \bold{b} \eta^{c} \wedge \eta^{a} +  \bold{\tilde{c}} \eta^{a} \wedge \eta^{b} )\nonumber \\
&+ &\eta^{\parallel} \wedge ( \eta^b - v_{\parallel} \eta^t) + \eta^{\perp} \wedge ( \eta^c - v_{\perp} \eta^t) - v_{\perp} \eta^{\xi} \wedge \eta^a.  \nonumber \\
\label{eq:dth1_expr}
\end{eqnarray}
The condition (\romannumeral2) can thus be verified. Moreover, acting $i_{\zeta}$ to both side of Eq.(\ref{eq:dth1_expr}), we see that $i_{\zeta} d \vartheta_1 \in \mathcal{N}^*_{\perp}$ and hence that the gyro-vector $\zeta$ play exactly the role of $Y_0$ in Eqs.~(\ref{eq:oscillating_s1}) and (\ref{eq:oscillating_sn}), making the gauge functions $\tilde{S}_n$ available through direct integrations over the gyro-angle $\xi$. The remaining task is then to recursively solve the generating vectors $G_n$'s according to Eqs.~(\ref{eq:sigl_G1_det}) or (\ref{eq:sigl_Gn_det}) after the symmetric part of both $\vartheta^{(n)}_n$ and $\vartheta^{(n+1)}_{n+1}$ are prudently chosen.

However, the influence of choosing different $\bar{\vartheta}_n$, the symmetric part of $\vartheta^{(n)}_n$, remains to be clarified. As mentioned in Sec.~\ref{sec:general}, all the components of $\bar{\vartheta}_n$, except for $i_{Y_0} \bar{\vartheta}_n$ which is determined together with the gauge function $\tilde{S}_n$, can be chosen as any functions independent on $\xi$, even those of little physical relevance, without affecting the solvability of each generating vector $G_n$. As to the current case, the $\bar{\vartheta}_{1}$ are free only in $\mathcal{N}^*_{\perp}$  due to the singularity of $d \vartheta_0$ and can be expressed as follows,
\begin{eqnarray}
\bar{\vartheta_1}= \boldsymbol{\bar{\vartheta}_{\bold{1,x}}} \cdot ( \bold{\tilde{a}} \eta^a +\bold{\tilde{c}} \eta^c )+ \langle \vartheta_1 \rangle ,
\label{eq:th11_free_part}
\end{eqnarray}
where the matrix $\boldsymbol{\bar{\vartheta}_{\bold{1x}}}$ consists of three arbitrary functions independent of $\xi$ and satisfies the constraint $\boldsymbol{\bar{\vartheta}_{\bold{1x}}} \cdot \bold{b} =0$. The singularity of $d \vartheta_0$ is compensated by $d \vartheta_1$ except along $\zeta$, and therefore the $\bar{\vartheta}_{n}$ for $n \ge 2$ are free in the total $T^*\mathcal{M}$ except along $\eta^{\xi}$ and can be constructed as
\begin{eqnarray}
\bar{\vartheta}_n=\bar{\vartheta}_{n,t} \eta^t +  \boldsymbol{\bar{\vartheta}_{\bold{n,x}}}\cdot ( \bold{\tilde{a}} \eta^a + \bold{b} \eta^b +\bold{\tilde{c}} \eta^c ) \nonumber \\
+ \bar{\vartheta}_{n,\parallel} \eta^{\parallel} + \bar{\vartheta}_{n,\perp} \eta^{\perp} + \langle i_{\zeta}\vartheta^{(n-1)}_{n} \rangle \eta^{\xi},
\label{eq:thnn_free_part}
\end{eqnarray}
with the $\bar{\vartheta}_{n,t}$, $\bar{\vartheta}_{n,\parallel}$, $\bar{\vartheta}_{n,\perp}$ and the elements of the matrix $\boldsymbol{\bar{\vartheta}_{\bold{n,x}}}$ chosen as arbitrary functions independent on $\xi$. Suppose we have solved all the generating vectors $G_n$, $n \le \mathcal{N}$ with the above $\bar{\vartheta}_n$'s so as to map the original $\Theta$ defined in Eqs.~(\ref{eq:norm_xi_eps}) into the $\Theta^{(N)}$ as follows,
\begin{eqnarray}
\Theta^{(N)} &=& d S^{(N)} + \vartheta^{(N)}_0 + \epsilon  \vartheta^{(N)}_1 + \cdots  \epsilon^N  \vartheta^{(N)}_N +  \cdots \nonumber \\
&=& d S^{(N)}  + \sum^{N}_{k=0} \epsilon^k (d \tilde{S}_k + \bar{\vartheta}_k) + o(\epsilon^{N+1}).
\end{eqnarray}
Denote the vortex vector of $\Theta^{(N)}$ by $\bar{X}$. Then, by construction, the scalar function $\bar{\mu}$,
\begin{eqnarray}
\bar{\mu} \equiv \sum_{k=0}^{N} \epsilon^{k} i_{\zeta} \bar{\vartheta}_k,
\label{eq:barmu_def}
\end{eqnarray}
is an $N$th-order invariant of $\bar{X}$, i.e.,
\begin{eqnarray}
L_{\bar{X}} \bar{\mu}= \langle \bar{X}, d \bar{\mu} \rangle = o(\epsilon^{N+1}).
\end{eqnarray}
As is seen from Eq.~(\ref{eq:thnn_free_part}) , the $n$th-order term $i_{\zeta} \bar{\vartheta}_n$ in the formal series of $\bar{\mu}$, though not directly chosen by us, generally depends on all the choices we have made for the preceding $i_{Y_{\alpha}}  \bar{\vartheta}_k$ with $\alpha \ne \xi$ and $k < n$. It follows that $\bar{\mu}$ is usually not unique. But this has no physical consequence as the flow of $\bar{X}$ is a non-physical one with the components of $\bar{\vartheta}_n$ freely chosen. What we expect to be unique is the $N$th-order scalar invariant of $X$, i.e. the formal expansion of magnetic moment $\mu$, which corresponds to the near-symmetry of $\Theta$ along the gyro-vector $\zeta$ and is obtained by the pull-back transformations from $\bar{\mu}$, namely, by the following relation,
\begin{eqnarray}
\mu &\equiv&  (\mathcal{F}^{-1})^* \bar{\mu}= exp(\epsilon L_{ G_1}) \cdots exp(\epsilon^N L_{ G_N}) \bar{\mu},
\label{eq:mu_def}
\end{eqnarray}
where $\mathcal{F}$ is defined in Eq.~(\ref{eq:comps_diffmorp}). Taking the exterior derivative $d$ of the both sides of the above equation (and using the commutativity between $d$ and induction maps), we see that
\begin{eqnarray}
d\mu &=& (\mathcal{F}^{-1})^*  \sum_{k=0}^{N} \epsilon^{k}  d i_{\zeta} \bar{\vartheta}_k \nonumber \\ &=& (\mathcal{F}^{-1})^*  \sum_{k=0}^{N} \epsilon^{k}  (-i_{\zeta} d ) \bar{\vartheta}_k \nonumber \\ &=& - (\mathcal{F}^{-1})^* i_{\zeta} d \Theta^{(N)} + o(\epsilon^{N+1}) \nonumber \\
&=&- (\mathcal{F}^{-1})^* i_{\zeta} d\mathcal{F}^* \Theta + o(\epsilon^{N+1}),
\label{eq:dmu}
\end{eqnarray}
where use has been made of $L_{\zeta} \bar{\vartheta}_k=0$. If different $\bar{\vartheta}_n$ were chosen, we obtain a different series of generating vectors $G_n^{'}$, a different composite diffeomorphism $\mathcal{F}^{'}$ and finally the $d \mu^{'}$ expressed similarly as in Eq.~(\ref{eq:dmu}) but with $\mathcal{F}^{'}$. As is proved in Ref.~\onlinecite{kruskal1962asymptotic} and further elaborated in Ref.~\onlinecite{burby2013auto}, the diffeomorphisms $\mathcal{F}$ and $\mathcal{F}^{'}$ differ each other only by a diffeomorphism $\Delta \mathcal{F} \equiv \mathcal{F}^{'} \circ \mathcal{F}^{-1}$ that commutes with the flow of $\zeta$. It follows immediately that $i_{\zeta}$ commutes with the induction maps of $\Delta \mathcal{F}$ (for the $\Delta \mathcal{F}$ is the flow of some $\Delta G$ that satisfies $[ \Delta G, \zeta ] = 0$) and that
\begin{eqnarray}
d\mu^{'}&=& - (\mathcal{F}^{'-1})^* i_{\zeta} d\mathcal{F'}^* \Theta + o(\epsilon^{N+1}) \nonumber \\
&=&- (\mathcal{F}^{-1})^* (\Delta \mathcal{F}^{-1})^*i_{\zeta} (\Delta\mathcal{F})^* d\mathcal{F}^* \Theta + o(\epsilon^{N+1})\nonumber \\
&=&- (\mathcal{F}^{-1})^* i_{\zeta} d\mathcal{F}^* \Theta + o(\epsilon^{N+1}) \nonumber \\
&=& d\mu + o(\epsilon^{N+1}).
\label{eq:dmu_prime}
\end{eqnarray}
Therefore, the adiabatic series $\mu$ obtained from choosing different $\bar{\vartheta}_n$ must be unique up to an inconsequential constant. In addition, as the $\mathcal{F}$ is a near-identity map, the lowest-order non-vanishing term of the $\bar{\mu}$, i.e. $i_{\zeta} \bar{\vartheta_2}$, is also unique and identical to that of the $\mu$. This can be proved by directly calculating the components of $G_1$ in $\mathcal{N}_{\perp}$ with the term $\boldsymbol{\bar{\vartheta}_{\bold{1x}}}$ freely chosen. The $i_{\zeta}\vartheta^{(2)}_2$ we obtained subsequently from Eq.~(\ref{subeq:sigl_G1_det_2}) must take the following form,
\begin{eqnarray}
i_{\zeta}\vartheta^{(2)}_2 = \frac{v^2_{\perp}}{2|\bold{B_0}|} - \frac{v_{\perp}}{2 |\bold{B_0}|} \boldsymbol{\bar{\vartheta}_{\bold{1x}}} \cdot \bold{\tilde{c}} \equiv i_{\zeta} \bar{\vartheta_2} + \frac{\partial \tilde{S}_2}{\partial \xi}.
\end{eqnarray}
It's seen that $i_{\zeta} \bar{\vartheta_2}$ is just the lowest-order magnetic moment unaffected by the free choices of $\boldsymbol{\bar{\vartheta}_{\bold{1x}}}$. In what follows we will replace $\bar{\mu}/\epsilon^2$ with $\bar{\mu}$, and $\mu/\epsilon^2$ with $\mu$, as both of the Eqs.(\ref{eq:barmu_def}) and (\ref{eq:mu_def}) start with the $o(\epsilon^2)$ terms.

Setting all the free components of $\bar{\vartheta}_n$ and $ G_n^{\xi}$ to zero, the $\Theta^{(n)}$ for $n \ge 2$ take the form of the Eq.~(\ref{eq:THn_final}) below, whose vortex vector, denoted by $X^{(n)}$, is given in Eq.~(\ref{eq:vortex_vector_final}) for the general $\bar{\mu}$. Although the representation of the vortex vector $X^{(n)}$ can be greatly simplified in the natural basis of $\{t, \bold{x}, v_{\parallel}, \bar{\mu}, \xi \}$, we stick to the coordinates $\{t, \bold{x}, v_{\parallel}, v_{\perp}, \xi \}$ for consistency. In formulating our results we have made use of the following definitions,
\begin{subequations}
\begin{eqnarray}
\bold{\bar{B}_1} &\equiv& \langle \bold{\tilde{B}_1} \rangle = \bold{B_1} + v_{\parallel} \nabla \times \bold{b} + \nabla \times \bold{u_E},  \\
\bold{\bar{E}_1} &\equiv& \langle \bold{\tilde{E}_1} \rangle =  \bold{E_1} - \frac{1}{2} \nabla (\bold{u_E} \cdot \bold{u_E}) + \bold{u_E} \times \bold{\bar{B}_1}, \\
\bold{\tilde{H}_1} &\equiv& \bold{\bar{B}_1} + \frac{1}{2} v_{\perp} \nabla \times \bold{\tilde{c}}, \\
\bold{\tilde{D}_1} &\equiv& \bold{\bar{E}_1} + \frac{1}{2} v_{\perp}\left( \bold{u_E} \times ( \nabla \times \bold{\tilde{c}}) - \nabla(\bold{u_E} \cdot \bold{\tilde{c}})\right).
\end{eqnarray}
\end{subequations}
The generating vectors $G_1$ and $G_2$ calculated respectively from Eqs.~(\ref{eq:sigl_G1_det}) and (\ref{eq:sigl_Gn_det}) are presented in Eqs.~(\ref{eq:Gn_res}) below.  For simplicity, both the $\bar{\mu}$ and $\mu$ are provided up to first order in Eqs.~(\ref{eq:barmu_res}) and (\ref{eq:mu_res}) respectively. In respond to a recent criticism\cite{zheng2023modifi} against the ordering consistency, we briefly mention that the curvature- and gradient-drift are contained in $X^{(n)}$ in the quantities $\bold{\bar{B}_1}$ and $\nabla \bar{\mu}$ respectively, both of them being consistently terms of $o(\epsilon^1)$ as can be seen in the components along $\partial / \partial \bold{x}$ in Eq.~(\ref{eq:vortex_vector_final}).

\section{\label{sec:concl} Conclusions and discussions}
The HLTP approach to the guiding-center motion, despite its complexities in terms of the mathematical apparatus involved, relies essentially on the simple linear transformation from the vector space of generating vectors ($T\mathcal{M}$) to that of the one-forms ($T^* \mathcal{M}$). However high the order of transformation is considered, such linear transformation is established by the lowest-order terms of the two-form arising from the exterior derivative of the PCOF in question, and is bound to be degenerate due to the existence of the null vectors, leading to the fact that the asymmetry in at least one component of the PCOF cannot be actually eliminated by HLTP transformation but has to be absorbed into an exact one-form generated by the gauge function. Therefore, the scheme of solving the generating vectors for HLTP approach could vary significantly depending on if the number of the null vectors is as minimal as possible, i.e., exactly one. If so, the null vector $Y_0$ (after its component in $\partial / \partial t$ is scaled to unity) is exactly the vortex vector of the zero-order term $\vartheta_0$ of the PCOF, and the determining equation of the generating vectors is then solved by firstly obtaining the gauge function through the partial differential equation along $Y_0$, and secondly inverting the linear transformation which, defined by $d \vartheta_0$ as in Eq.~(\ref{eq:nth_difference}) with inherent singularity, is invertible only when restricted into the subspaces $\mathcal{N}_{\perp} \in T\mathcal{M}$ and $\mathcal{N}_{\perp}^* \in T^*\mathcal{M}$. In such case, the feasibility of HLTP approach is theoretically guaranteed to arbitrary order provided only the assumption that the $\vartheta_0$ is symmetic along some vector $\zeta$, i.e. that $L_{\zeta} \vartheta_0=0$. In the case where $d \vartheta$ has multiple null vectors, however, two extra conditions must be satisfied by the first-order term $\vartheta_1$ of the PCOF in compensation for the singularity of the $\vartheta_0$: one is for ensuring the asymmetry along $\zeta$ does not exist in the subspace $\mathcal{N}^*$ as such asymmetry will be preserved in the succeeding PCOF, and the other indicates that $d \vartheta_1$ defines another linear transformation between the subspaces $\mathcal{N}$ and $\mathcal{N}^*$ where $d \vartheta_0$ fails to establish a useful relation between the generating vectors and one-forms. The general determining equations of the generating vectors for both cases have been provided in a component-wise manner, whereas the practicality of actually solving these equations depends on whether or not the integral curve of the null vector $Y_0$ can be analytically found so as to obtain the gauge functions that constitute the pure oscillating part of the generating vectors.

The guiding-center motion naturally fits into what we call the singular perturbation motion, as the leading term of its corresponding PCOF consist of only the equilibrium potential fields. Instead of seeking the preparatory transformations which convert the singular leading term to a non-singular one, we carry out the HLTP transformation following closely the general procedure in Sec.~\ref{sec:case2} that addresses directly the singularity of the leading term. In doing so, we have developed a staggered scheme of performing the HLTP approach, whereby the gyro-angle dependencies in the PCOF of each order is eliminated by the generating vectors of two adjoining order. The theoretical benefit of this novel scheme is first of all manifest in the gauge functions, which are obtained from direct integration over the gyro-angle rather than from approximate solution of partial differential equation, thereby leaving no errors in the generating vector of each order. Secondly, the adoption of the active point of view has enabled us to carry out this HLTP scheme in the non-perturbative coordinates systems $\{ t, \bold{x}, v_{\|}, v_{\perp},\xi \}$ and to circumvent the subtle problems such as where a physical quantity is evaluated at, hence the convenience in expressing the fields, one-forms, etc. Moreover, the perturbation quantities, such as those arising from the spatial non-uniformity of the equilibrium fields and those from the low-amplitude electromagnetic fluctuation, can be considered in a united framework retaining the effects of strong $E \times B$ shearing. Although little is mentioned throughout this work about how the systems of Maxwell's equations can be coupled with the gyrokinetic Vlasov equation obtained from Eq.~(\ref{eq:vortex_vector_final}), this problem can be addressed in the collisionless regime where the distribution function is itself an invariant of the vortex vector. Whenever the non-physical distribution function $\bar{f}$ is obtained from the numeric solution of $\langle X^{(n)}, d \bar{f}\rangle =0$ with the initial condition at some time point $t_0$, the particle distribution function $f$ is obtained immediately by the pull-back transformation as in Eq.~(\ref{eq:mu_def}), and the densities of charge and of current obtained by integrating $f$ in the coordinates $\{ t, \bold{x}, v_{\|}, v_{\perp},\xi \}$. The resulting gyrokinetic Vlasov-Maxwell system is thus suitable to the full-f gyrokinetic simulation as the equilibrium and perturbed part of any quantity is not treated separately throughout the HLTP derivation. The main results of this work can be validated by comparison with various previous work; for example, the adiabatic series of magnetic moment Eq.~(\ref{eq:mu_res}) can be compared with those provided in Ref.~\onlinecite{weyssow1986hamil,littlejohn1981hf}. Nevertheless, further validation is needed and will be carried out in future work adopting realistic equiilbrium magnetic field.

\begin{acknowledgments}
This work was supported by the National Natural Science Foundation of China with grant No.~12275071. It was also partially supported by Innovation Program of Southwestern Institute of Physics (Grants No.~202301XWCX001) and Science and technology project of Sichuan Province with grant No.~2023ZYD0016.
\end{acknowledgments}
\begin{widetext}
\begin{flalign}
&\Theta^{(n)}= dS^{(n)}+\bold{A}_{\bold{0}} \cdot d \bold{x}-\phi_0 d t+\epsilon\left(\left(\bold{A}_{\bold{1}}+v_{\|} \bold{b}+\bold{u}_{\bold{E}}\right) \cdot d \bold{x}-(\frac{1}{2} (v_{\|}^2+ v_{\perp}^2+ \bold{u}_{\bold{E}} \cdot \bold{u}_{\bold{E}} )+\phi_1) d t\right) & \nonumber \\
& \quad \quad \quad +\epsilon^2 \bar{\mu} d \xi + \sum_{k=3}^n \epsilon^k d \tilde{S}_k+o\left(\epsilon^{n+1}\right),&
 \label{eq:THn_final}
\end{flalign}
\begin{flalign}
&X^{(n)}  =\frac{\partial}{\partial t}+\bold{u_E} \cdot \frac{\partial}{\partial \bold{x}}+\frac{1}{\left(\bold{B}_{\bold{0}}+\epsilon \bar{\bold{B}}_{\bold{1}}\right) \cdot \bold{b}}\bigg[\Big((v_{\|}-\frac{\partial \bar{\mu} / \partial v_{\|}}{\partial \bar{\mu} / \partial v_{\perp}} v_{\perp})(\bold{B}_{\bold{0}}+\epsilon \bar{\bold{B}}_{\bold{1}})+\epsilon(\bar{\bold{E}}_{\bold{1}}+\frac{v_{\perp} \nabla \bar{\mu}}{\partial \bar{\mu} / \partial v_{\perp}}) \times \bold{b}\Big) \cdot \frac{\partial}{\partial \bold{x}}&\nonumber\\
& \quad \quad \quad+(\bold{B}_{\bold{0}}+\epsilon \bar{\bold{B}}_{\bold{1}}) \cdot\Big(\bar{\bold{E}}_{\bold{1}}+\frac{v_{\perp}}{\partial \bar{\mu} / \partial v_{\perp}} \bold{\nabla} \bar{\bold{\mu}}\Big) \frac{\partial}{\partial v_{\|}}-\frac{1}{\partial \bar{\mu} / \partial v_{\perp}}\Big((\bold{B}_{\bold{0}}+\epsilon \bar{\bold{B}}_{\bold{1}}) \cdot(\frac{\partial \bar{\mu}}{\partial v_{\|}} \bar{\bold{E}}_{\bold{1}}+v_{\|} \nabla \bar{\mu})+\epsilon \bar{\bold{E}}_{\bold{1}} \cdot(\bold{b} \times \nabla \bar{\mu})&\nonumber \\
& \quad \quad \quad+(\partial \bar{\mu} / \partial t+\bold{u}_{\bold{E}} \cdot \nabla \bar{\mu})(\bold{B}_{\bold{0}}+\epsilon \bar{\bold{B}}_{\bold{1}}) \cdot \bold{b}\Big) \frac{\partial}{\partial v_{\perp}}\bigg]+\frac{1}{\epsilon} \frac{v_{\perp}}{\partial \bar{\mu} / \partial v_{\perp}} \frac{\partial}{\partial \xi}+\mathrm{o}\left(\epsilon^n\right),& \label{eq:vortex_vector_final}
\end{flalign}
\begin{flalign}
 &G_1 = - \frac{v_{\perp}}{ \vert \bold{B_0} \vert} Y_a + \frac{v_{\perp}}{ \vert \bold{B_0} \vert} \bold{\tilde{H}_1} \cdot \bold{\tilde{c}} Y_{\parallel} - \frac{1}{ \vert \bold{B_0} \vert} \left( v_{\parallel} \bold{\tilde{H}_1} \cdot \bold{\tilde{c}} + \bold{\tilde{D}_1} \cdot \bold{\tilde{a}}\right) Y_{\perp}, &
\end{flalign}
\begin{subequations}
\label{eq:Gn_res}
\begin{flalign}
& G_2= \frac{1}{v_{\perp}} \bigg( -\frac{ \partial \tilde{S}_3}{\partial v_{\perp}}  Y_t +( v_{\perp}  \frac{\partial \tilde{S}_3}{v_{\parallel} }- v_{\parallel} \frac{\partial \tilde{S}_3}{\partial v_{\perp}} )Y_b -  ( v_{\perp} \bold{b} \cdot \frac{ \partial \tilde{S}_3}{\partial \bold{x}}+ \bold{\bar{E}_1}\cdot \bold{b}   \frac{\partial \tilde{S}_3}{\partial v_{\perp}} ) Y_{\parallel} +( \frac{\partial  \tilde{S}_3}{\partial t}  + \bold{u_E} \cdot \frac{\partial  \tilde{S}_3}{\partial \bold{x}}  + \bold{\bar{E}_1}\cdot \bold{b} \frac{\partial \tilde{S}_3}{\partial v_{\parallel}} + v_{\parallel} \bold{b} \cdot \frac{\partial \tilde{S}_3}{\partial \bold{x}} ) Y_{\perp}  \bigg) &\nonumber \\
& \quad \quad \quad +\frac{1}{2\left|\bold{B}_{\bold{0}}\right|^2}\bigg(2 v_{\perp} \tilde{\bold{H}}_{\bold{1}} \cdot \bold{b}-v_{\|} \tilde{\bold{H}}_{\bold{1}} \cdot \tilde{\bold{c}}-\tilde{\bold{D}}_{\bold{1}} \cdot \tilde{\bold{a}}\bigg) Y_a+\frac{1}{2\left|\bold{B}_{\bold{0}}\right|^2}\bigg(\frac{v_{\perp}^3}{6} \kappa_c +\tilde{\bold{B}}_{\bold{1}} \cdot \tilde{\bold{c}}\left(v_{\|} \tilde{\bold{H}}_{\bold{1}} \cdot \tilde{\bold{c}}+\tilde{\bold{D}}_{\bold{1}} \cdot \tilde{\bold{a}}\right) -2 v_{\perp} \tilde{\bold{H}}_{\bold{1}} \cdot \tilde{\bold{c}} \tilde{\bold{H}}_{\bold{1}} \cdot \bold{b}\bigg) Y_{\|} &\nonumber\\
&\quad \quad \quad+\frac{1}{2 v_{\perp}\left|\bold{B}_{\bold{0}}\right|^2}\bigg(-\frac{v_{\perp}^3}{6} \kappa_E -\frac{v_{\perp}^3 v_{\|}}{6}\kappa_c   +\left(v_{\|} \tilde{\bold{H}}_{\bold{1}} \cdot \tilde{\bold{c}}+\tilde{\bold{D}}_{\bold{1}} \cdot \tilde{\bold{a}}\right)\left(2 v_{\perp} \tilde{\bold{H}}_{\bold{1}} \cdot \bold{b}-v_{\|} \tilde{\bold{B}}_{\bold{1}} \cdot \tilde{\bold{c}}-\tilde{\bold{E}}_{\bold{1}} \cdot \tilde{\bold{a}}\right)\bigg) Y_{\perp}, &
\end{flalign}
\begin{flalign}
&\kappa_{E}\equiv|\bold{B_0}|\tilde{\bold{a}}\cdot\nabla(\frac{\bold{u_E}\cdot\nabla\tilde{\bold{a}}\cdot\tilde{\bold{c}}-\tilde{\bold{a}}\cdot\nabla\bold{u_E}\cdot\tilde{\bold{c}}}{|\bold{B}_{0}|})+\tilde{\bold{c}}\cdot(\nabla\times\tilde{\bold{c}})(\tilde{\bold{a}}\cdot\nabla\bold{b}\cdot\bold{u_E} -\bold{u_E}\cdot\nabla\bold{b}\cdot\tilde{\bold{a}})-\bold{b}\cdot(\nabla\times\tilde{\bold{c}})(\bold{u_E}\cdot\nabla\tilde{\bold{a}}\cdot\tilde{\bold{c}} -\tilde{\bold{a}}\cdot\nabla\bold{u_E}\cdot\tilde{\bold{c}}), &
\end{flalign}
\begin{flalign}
&\kappa_c \equiv |\bold{B_0}|\tilde{\bold{a}}\cdot\nabla(\frac{\tilde{\bold{c}}\cdot(\nabla\times\tilde{\bold{c}})}{|\bold{B_0}|})+\tilde{\bold{c}}\cdot(\nabla\times\tilde{\bold{c}})\tilde{\bold{c}}\cdot(\nabla\times\bold{b})-\bold{b}\cdot(\nabla\times\tilde{\bold{c}})\tilde{\bold{c}}\cdot(\nabla\times\tilde{\bold{c}}),&
\end{flalign}
\begin{flalign}
&\tilde{S}_3=\frac{v_{\perp}^3}{3\left|\bold{B}_{\bold{0}}\right|^2}\left(\tilde{\bold{c}} \cdot \frac{\nabla\left|\bold{B}_{\bold{0}}\right|}{\left|\bold{B}_{\bold{0}}\right|}-\bold{b} \cdot(\nabla \times \tilde{\bold{a}})\right)-\frac{v_{\perp}}{\left|\bold{B}_{\bold{0}}\right|^2}\bigg(\bar{\bold{E}}_{\bold{1}} \cdot \tilde{\bold{c}}-v_{\|} \bar{\bold{B}}_{\bold{1}} \cdot \tilde{\bold{a}}\bigg) & \nonumber \\
& \quad \quad \quad +\frac{v_{\perp}^2}{8\left|\bold{B}_{\bold{0}}\right|^2}\bigg(\bold{c} \cdot \nabla \bold{u}_{\bold{E}} \cdot \bold{a}+\bold{a} \cdot \nabla \bold{u}_{\bold{E}} \cdot \bold{c}+v_{\|}(\nabla \times \bold{a}) \cdot \bold{a}-v_{\|}(\nabla \times \bold{c}) \cdot \bold{c}\bigg) \sin (2 \xi)& \nonumber \\
& \quad \quad \quad +\frac{v_{\perp}^2}{8\left|\bold{B}_{\bold{0}}\right|^2}\bigg(\bold{a}\cdot \nabla \bold{u_E}\cdot \bold{a} - \bold{c}\cdot \nabla \bold{u_E}\cdot \bold{c} - v_{\|} \left( \nabla \times \bold{a} \right) \cdot \bold{c} - v_{\|} \left( \nabla \times \bold{c} \right) \cdot \bold{a}\bigg) \cos (2 \xi), &
\end{flalign}
\end{subequations}
\begin{flalign}
&\bar{\mu}=\frac{v_\perp^2}{2|\bold{B_0}|}+\epsilon\frac{v_\perp^2}{4|\bold{B_0}|^2}\Big(v_\parallel(\nabla\times\bold{a})\cdot\bold{a}+v_\parallel(\nabla\times\bold{c})\cdot\bold{c}-2\bold{\bar{B}_1} \cdot \bold{b}+(\nabla \times \bold{u_E})\cdot \bold{b}+2\bold{u}_E\cdot\nabla\bold{a}\cdot\bold{c}\Big)+o(\epsilon^2),& \label{eq:barmu_res}
\end{flalign}
\begin{flalign}
&\mu=\frac{v_\perp^2}{2|\bold{B_0}|}+\epsilon\frac{v_\perp}{4|\bold{B_0}|^2}\Big(v_\perp v_\parallel(\nabla\times\bold{a})\cdot\bold{a}+v_\perp v_\parallel(\nabla\times\bold{c})\cdot\bold{c}-2v_\perp \bold{\bar{B}_1} \cdot \bold{b}  +v_{\perp} (\nabla \times \bold{u_E})\cdot \bold{b} +2v_\perp\bold{u}_E\cdot\nabla\bold{a}\cdot\bold{c} & \nonumber \\
&\quad \quad \quad +2v_\perp^2\tilde{\bold{a}}\cdot\frac{\nabla|\bold{B_0}|}{|\bold{B_0}|}-4v_\parallel\bold{\tilde{H}_1}\cdot\tilde{\bold{c}}-4\bold{\tilde{D}_1}\cdot\tilde{\bold{a}}\Big)+o(\epsilon^2).& \label{eq:mu_res}
\end{flalign}

\end{widetext}

\bibliography{GF_in_LT}
\end{document}